\documentclass[11pt,a4paper]{article}
\usepackage{jheppub}
\usepackage{graphicx,amsmath,epsfig}
\def\beq{\begin{equation}} 
\def\eeq{\end{equation}} 
 
\def\beqn{\begin{eqnarray}} 
\def\eeqn{\end{eqnarray}} 

\def\slashit#1{\slash \mkern-12mu#1}

\newcommand{\mph}[1]{ m_{#1,{\rm phys}}  }
\newcommand{\physd}[2]{ {#1}_{#2,{\rm phys}} }
\newcommand{\physs}[1]{{#1}_{\rm phys} }

\def\simge{\mathrel{%
   \rlap{\raise 0.511ex \hbox{$>$}}{\lower 0.511ex \hbox{$\sim$}}}}

\def\simle{\mathrel{
   \rlap{\raise 0.511ex \hbox{$<$}}{\lower 0.511ex \hbox{$\sim$}}}}

\newcount\minutes 
\newcount\scratch 
 
\def\timestamp{%
\scratch=\time 
\divide\scratch by 60 
\edef\hours{\the\scratch} 
\multiply\scratch by 60 
\minutes=\time 
\advance\minutes by -\scratch 
---$\,$\hours:\null 
\ifnum\minutes< 10 0\fi 
\the\minutes} 

\title{The other Higgses, at resonance, in the Lee-Wick extension of the Standard Model}

\author[a]{Terrance Figy}
\author[b]{Roman Zwicky}
\affiliation[a]{TH Division, CERN, CH-1211 Geneva 23, Switzerland}
\affiliation[b]{Southampton University, Highfield, SO17 1BJ Southampton, UK}
\emailAdd{terrance.maynard.figy@cern.ch}
\emailAdd{r.zwicky@soton.ac.uk}

\abstract{Within the framework of the Lee Wick 
  Standard Model (LWSM) we investigate 
  Higgs pair production $gg \to h_0 h_0$, $gg \to h_0 \tilde p_0$ 
   and top pair production $gg \to \bar tt$ at the Large Hadron Collider (LHC), 
   where the neutral particles
   from the Higgs sector  ($h_0$, $\tilde h_0$ and $\tilde p_0$)
   appear as possible resonant intermediate states.
   We investigate the signal  $gg \to h_0 h_0 \to \bar b b \gamma \gamma$
   and we find that the LW Higgs, depending on its mass-range, can be seen not long after the LHC upgrade in 2012. 
   More precisely this happens when the new LW Higgs states are below the top pair threshold.
    In $gg \to \bar tt$ the LW states, due to the wrong-sign propagator 
   and negative width, lead to a dip-peak structure instead of the usual peak-dip structure which 
   gives a characteristic signal especially for low-lying LW Higgs states.
  We comment on the LWSM and the forward-backward asymmetry in view of 
  the measurement at the TeVatron.
  Furthermore, we present a technique which reduces the hyperbolic diagonalization 
  to standard diagonalization methods. We clarify issues of spurious 
  phases in the Yukawa sector. }

\arxivnumber{YYMM.NNNN}
\subheader{CERN-PH-TH/2011-184\\
          CP-3/2011-25}

\begin{document}
\maketitle
\today \timestamp \hfill 
\vfill

\section{Introduction}
\label{sec:intro}
\subsection{The Lee-Wick Standard Model}
The investigation of the mechanism of electroweak symmetry breaking (EWSB), responsible
for the generation of fermion and gauge boson masses,
is one of the primary tasks of the Large Hadron Collider (LHC) at CERN. 
The scalar Higgs particle realizes this mechanism in the Standard Model (SM), in a rather 
efficient way, at the expense of divergences quadratic in the cut-off. 
The latter fact, known as the hierarchy problem, is taken as an  indication of the 
incompleteness of the SM and is at the heart of many models beyond the SM (BSM).
An example of which is the  Lee-Wick SM (LWSM) \cite{Grinstein:2007mp} 
where ideas to  soften ultraviolet (UV)  divergences in QED  
from the seventies \cite{negmetric,finiteqed}  were extended to chiral fermions and non-abelian gauge theories \cite{Grinstein:2007mp}.
Most importantly it was shown that the LWSM is renormalizable 
and free from quadratic divergences  \cite{Grinstein:2007mp} thus joining the list of 
models addressing the hierarchy problem successfully.
In LW field theories higher derivative (HD) terms are added and terms quadratic in the fields 
are resummed into the propagator rather than treated as perturbations,
ameliorating the UV behavior of perturbation theory. 
This results in additional poles in the  propagators 
for which  auxiliary fields (AF) can be introduced to cast the theory 
in terms of interactions with mass dimension no greater than four\footnote{It is amusing 
that in the AF-formalism the LWSM seems fine tuned with respect to the hierarchy problem whereas this is not the case in the HD-formalism as a single term is added in each sector.}. 
The additional fields are interpreted as  LW partner states and do have the
wrong-sign propagator, aka Pauli-Villars regulators.
The key idea of Lee and Wick is that the LW ghost particles never appear as asymptotic
states in detecters, nowadays reminiscent of the Faddev-Poppov ghosts in non-abelian gauge field theories.
The connected issues of unitarity and causality which were debated 
in the seventies, e.g. the Erice lectures \cite{Lee-Erice,Coleman-Erice},
and reconsidered recently in \cite{Grinstein:2008bg}.
Most notably the width becomes negative and requires a 
deformation of the contour to avoid new cuts \cite{S-gang} which assure no new asymptotic states.
The status of LW field theories is that there are no known counterexamples  to unitarity in
perturbation theory up to today and that causality can be violated but 
only at distances below $M_{\rm LW}^{-1}$.  It has been suggested that the 
violation of causality can be tested at the LHC \cite{Alvarez:2009af}.
The usual non-perturbative formulation via the path-integral seems
difficult \cite{BoulawareGross} but recently a restrictive path-integral was proposed 
where the contour prescription can be derived \cite{tonder,tonder1}. 

Further conceptual issues of phenomenological  nature have been investigated such as the behaviour 
at high temperature \cite{Fornal:2009xc}, unitarity of massive LW vector boson scattering \cite{massive},
the compatibility of the see-saw and the absence of quadratic divergences \cite{Espinosa:2007ny}, 
the running of couplings \cite{Grinstein:2008qq}, UV-properties of LW field theories \cite{Espinosa:2011js},
even higher derivative LW field theories \cite{Carone:2008iw,Carone:2009it} and
LW fields and gravity \cite{Rodigast:2009en}. The cosmology of LW field theories has been investigated in  \cite{Cai:2008qw}.
Phenomenological studies include LHC and linear collider signals of LW gauge bosons \cite{RizzoLHC,RizzoILC}, flavour changing 
neutral currents \cite{DW}, 
electroweak precision observables (EWPO) have been investigated in 
\cite{Underwood:2008cr} and \cite{Chivukula:2010nw} where
gauge boson and fermion masses are found to be constrained 
up to  a few TeV. 

\subsection{The Higgs-sector of the Lee-Wick SM}
The LW Higgs sector has been investigated 
in \cite{Krauss:2007bz,Carone:2009nu,Cacciapaglia:2009ky,Alvarez:2011ah}.
The neutral part consists of the CP-even $h_0$, $\tilde h_0$, which are the SM-like 
and the LW-like Higgs boson, and the CP-odd LW-like scalar $\tilde p_0$. 
The SM as well as the LW Higgs sectors are not easy to constrain, neither
indirectly through loops nor directly through signals. 
First for large Higgs masses the latter
enters only logarithmically, rather than quadratically,  at one-loop \cite{Veltman}. 
Second the Higgs couples via Yukawa terms to fermions 
and is therefore highly suppressed  in di-lepton signals $h \to l^+l^-\;\;$\footnote{This is why, in our opinion,  
the LW Higgs is not a candidate for the $Wjj$-excess at the TeVatron \cite{Aaltonen:2011mk} as it should
already have been seen in $Wll$-signal or $Wbb$-signal.}.

A salient feature of the LWSM, at least in its minimal version \cite{Grinstein:2007mp}, 
is that there's roughly a single new parameter per sector. It's the mass in the HD formalism 
which predicts all masses and couplings in the language of the AF formalism.
In this respect the LWSM resembles so-called sequential SM extensions.
The aim of this paper is to investigate the effect of a low lying Higgs sector, as 
a function of this single new parameter and the Higgs mass. 
We focus on channels, accessible at the LHC, where the additional Higgs  appear 
as  intermediate states at or close to resonance. 
\begin{itemize}
\item \emph{Higgs boson pair production}  is beyond reach at the  LHC in the SM \cite{Baur:2003gp}.
In extensions of the SM its a different quest as particles, with appropriate couplings
and masses above the two Higgs threshold, can enhance the cross section by orders
of magnitude without contradicting current constraints \footnote{A well-known example is  minimal supersymmetric SM (MSSM) \cite{Plehn:1996wb}. In fact the LWSM Higgs sector particle content corresponds to a type-II two Higgs-doublet model  with a new LW Higgs mass scale as a  single new parameter with $\tan \beta =1$. The masses of the different Higgs particles are discussed in section \ref{sec:lwsm}.}.
We consider   
$gg  \to h_0 h_0$ and $gg \to \tilde p_0 h_0$. We find that the  cross section of the latter can be enhanced by roughly three orders of magnitude with respect to the SM for a sizable range of masses. That is to say if the LW Higgs is above the SM-like Higgs pair threshold and not 
to far above the top pair threshold, $2 m_{h_0} < m_{\tilde h_0} \simle  1.5 (2 m_t) $. 
If the latter bound is approached top pair production becomes the main channel:

\item \emph{top pair production through gluon fusion} does not suffer from low cross sections
and has already been observed at colliders. The cross section of the invariant mass
of the top-pair $M_{tt}$ has been identified as an attractive observable to see resonance
effects through interference with the QCD-part a long time ago e.g. \cite{Dicus:1994bm}. 
LW field theories have a very different pattern in that the wrong-sign 
propagator and width lead to a dip-peak rather than a peak-dip structure in 
the spectrum.  It should be added that such effects 
can and do also appear in strongly coupled theories such as low energy QCD as discussed in section \ref{sec:toppair}.
\end{itemize}

The paper is organized as follows: In section~\ref{sec:lwsm} we give an overview of the Higgs and quark
sectors within the LWSM. In sections~\ref{sec:hpair}  and \ref{sec:toppair}  we discuss Higgs pair 
and top pair production from a theory point of view. In section \ref{sec:tFBA} we comment
on the top forward-backward asymmetry in view of the current TeVatron  results. In section 
\ref{sec:numerical} we present plots. In section \ref{sec:signal} 
we investigate the signal $gg \to h_0 h_0 \to \bar b b \gamma \gamma$.
In section \ref{sec:end} we conclude. In appendices \ref{app:gghh} and \ref{app:ttresults} we present further details of amplitudes for Higgs pair and top pair production, respectively.
In appendix \ref{app:mass}  a method that reduces the hyperbolic diagonalization 
to standard techniques is presented. In appendices \ref{app:masssumrules} and \ref{app:spurious}
we present tree-level mass sum rules. Further, we clarify the issue of spurious phases versus 
CP-violating phases in the fermion mass matrices.

\section{The Lee Wick Standard Model} 
\label{sec:lwsm}
We shall discuss the Higgs and Yukawa sectors directly in the auxiliary field formalism 
and refer the interested reader to \cite{Grinstein:2007mp} for the connection with
the higher derivative formalism.

\subsection{Higgs sector}
\label{sec:HiggsLW}

The Lagrangian of the Higgs sector in the auxiliary field formalism assumes the 
following form \cite{Grinstein:2007mp}:
\begin{eqnarray}
\mathcal{L} = (\hat D_{\mu}H)^{\dagger} (\hat D^{\mu}H)
             -(\hat D_{\mu}\tilde{H})^{\dagger} (\hat D^{\mu}\tilde{H}) 
             + M_{H}^{2} \tilde{H}^{\dagger} \tilde{H} - V(H-\tilde{H}) \;,
\end{eqnarray} 
where $\hat D_\mu = 
\partial_\mu +i ({\bf A}_\mu + {\bf \tilde  A}_\mu)$ 
with  ${\bf A_\mu} = g A_\mu^a T^a + g_2 W_\mu^a T^a
+ g_1 B_\mu\, Y$ for  SM gauge fields and analogously for 
the LW gauge boson for $\tilde {\bf A}_\mu$.  The Higgs potential 
is $V(H) = \lambda/4( H^\dagger H - v^2/2)^2$. The mass $M_H$ is 
the mass scale of the higher derivative LW mass scale.
In the unitary
gauge the two doublets are
\begin{equation}
\label{eq:unitarygauge}
H^\top = \big[0, (v+h_0)/\sqrt{2}\big]\,,\quad 
\tilde{H}^\top = 
\big[\tilde{h}_+,(\tilde{h}_0+i\tilde{p}_0)/\sqrt{2}\big]\, .
\end{equation} 
It is worthwhile to emphasize that, prior to mixing, the SM but not the LW 
CP-even neutral Higgs acquires a vacuum expectation value:
\begin{equation}
\langle h_0 \rangle  = v \;, \quad \langle \tilde h_0 \rangle = 0 \;.
\end{equation}
We note the  standard abuse of notation in not denoting the massless as well 
as the massive Higgs field by $h_0$. With \eqref{eq:unitarygauge} 
the mass Lagrangian assumes the following form:
\begin{eqnarray}
\label{eq:mh}
 \mathcal{L}_{\textrm{mass}} = -\frac{\lambda}{4} v^2 (h_0 - \tilde{h}_0)^2 
+ \frac{M_{H}^{2}}{2} (\tilde{h}_0 \tilde{h}_0 + \tilde{p}_0 \tilde{p}_0 + 2 \tilde{h}_{+} \tilde{h}_{-}) \, .
\end{eqnarray}
We note the mixing between the Higgs scalar and its LW--partner.
The neutral CP-even Higgs field can be diagonalized by a symplectic rotation:
\begin{eqnarray}
 \begin{pmatrix}
  h \\
  \tilde{h}
 \end{pmatrix} = 
\begin{pmatrix}
\cosh \phi_{h} & \sinh \phi_{h} \\
\sinh \phi_{h} & \cosh \phi_{h}
\end{pmatrix} 
\begin{pmatrix}
  h_{\textrm{phys}} \\
  \tilde{h}_{\textrm{phys}}
 \end{pmatrix} \, .
\end{eqnarray}
for which the masses of the Higgs sector are given by,
\begin{equation}
\label{eq:Htable}
\renewcommand{\arraystretch}{1.5}
\addtolength{\arraycolsep}{9pt}
\begin{array}{l | r | r | r | r  }
                     & h_0 & \tilde h_0 & \tilde p_0 & h_{\pm} \\[0.1cm] 
                     \hline
                       \text{CP} & \text{even} & \text{even} & \text{odd} & \text{none} \\[0.1cm]
\hline
 \frac{m^2_{\rm phys}}{M_H^2} 
 &  \frac{1}{2} \left(1 - \sqrt{1 - 2 v^2 \lambda/M_{H}^2} \right)  & 
  \frac{1}{2}   \left(1 + \sqrt{1 - 2 v^2 \lambda/M_{H}^2} \right)   &  1 & 1 
 \end{array}
\renewcommand{\arraystretch}{1}
\addtolength{\arraycolsep}{-9pt}
\end{equation}
and for completeness we have indicated the CP quantum numbers as well.
For obtaining Feynman rules in terms of the physical masses the 
following relations are useful \cite{Krauss:2007bz}:
\begin{eqnarray}
\label{eq:lambda}
 \lambda v^2 = \frac{2 m^{2}_{h_0 , \textrm{phys}}}{( 1 + r_{h_0}^2  )}  
                    \,, 
\qquad \quad r_{h_0} \equiv \frac{\mph{ h_0}} {\mph{\tilde h_0}}\,,
\end{eqnarray}
and
\begin{eqnarray}
\label{eq:s}
s_H &=& \cosh \phi_h  =  \frac{1}{( 1 - r_{h_0}^4  )^{1/2} }\;,
 \nonumber  \\
s_{H\!-\!\tilde H} &=& \cosh \phi_h  - \sinh \phi_h =
\frac{ 1 + r_{h_0}^2  }{( 1 - r_{h_0}^4  )^{1/2} } 
 \, .
\end{eqnarray}

\subsection{Yukawa Interactions}
In order to discuss the Yukawa terms, it is helpful to first discuss 
the fermions. We shall closely follow ref.~\cite{Krauss:2007bz}. 
However, we choose a slightly different basis for the fermions 
and refer the reader to appendix \ref{app:mass} where a method 
is outlined how the hyperbolic diagonalization can be performed
using standard tools.

The kinetic term of the AF Lagrangian is given by:
\begin{eqnarray}
 \mathcal{L} = \overline{\Psi^{t}} i \eta_{3} \hat{\slashit D} \Psi^{t} 
- \overline{\Psi^{t}_{R}} \mathcal{M}_{t} \eta_{3} \Psi^{t}_{L} - 
\overline{\Psi^{t}_{L}} \eta_{3} \mathcal{M}^{\dagger} \Psi^{t}_{R} \, ,
\end{eqnarray}
with 
\begin{eqnarray}
\label{eq:multiplet}
 \Psi_{L}^{t \top}=(T_{L}, \tilde{t}^{\prime}_{L} , \tilde{T}_{L}) \, , \quad \quad 
\Psi_{R}^{t \top} = (t_{R}, \tilde{t}_{R}, \tilde{T}^{\prime}_{R}) \, ,
\end{eqnarray}
where all capitalized components are part of an SU(2) doublet; e.g. $Q_L  = (T_L,B_L)^\top$.
It is noteworthy that a chiral fermion necessitates two chiral fermions which in turn form
a massive Dirac fermion. This becomes explicit in the basis chosen above 
 \begin{eqnarray}
 \label{eq:Mmatrix}
\mathcal{M}_{t} \eta_{3} = 
\begin{pmatrix}
 m_{t} & 0 &-m_{t}  \\
-m_{t} &  -M_{u} & m_{t} \\
0 & 0 & -M_{Q}  
\end{pmatrix}    \;,  \qquad  \eta_{3} = 
\begin{pmatrix}
 1 & 0 & 0  \\
0 & -1  & 0 \\
0 & 0 & -1 
\end{pmatrix}
\end{eqnarray}
which differs from the one in  \cite{Krauss:2007bz}. Note though that all physical masses
remain unchanged under change of basis. The mass matrix is diagonalized 
by symplectic rotations $S_L$ and $S_R$:
\begin{eqnarray}
 \Psi_{L(R), \textrm{phys}} = \eta_{3} S_{L(R)}^{\dagger} \eta_{3} \Psi_{L(R)} \,, \qquad   \mathcal{M}_{t, \textrm{phys}} \eta_{3}=  S_{R}^{\dagger} \mathcal{M}_{t} \eta_{3} S_{L} \;,
\end{eqnarray}
which leave the kinetic terms invariant by virtue of 
\begin{eqnarray}
\label{eq:SnSn}
 S_{L} \eta_{3} S_{L}^{\dagger} = \eta_{3} \quad {\rm and} \quad S_{R}
 \eta_{3} S_{R}^{\dagger} = \eta_{3} \, .
\end{eqnarray}

Now we may turn to the Yukawa sector for which we only write down the neutral Higgs part:
\begin{equation}
\label{eq:hqq}
 \mathcal{L}=-\frac{1}{v}(h_0 - \tilde{h}_0)\Big( \overline{\Psi_{R}^{t}} g_{t} \Psi_{L}^{t} + 
  \overline{\Psi_{L}^{t}} g_{t}^\dagger \Psi_{R}^{t}\Big)  -\frac{1}{v}(- i \tilde p_0) \Big( \overline{\Psi_{R}^{t}} g_{t} \Psi_{L}^{t} - 
  \overline{\Psi_{L}^{t}} g_{t}^\dagger \Psi_{R}^{t}\Big)     \;,
\end{equation}
where the $g$ matrix has non-diagonal entries which allow for transitions
between LW-generations and is given in the initial and physical basis by:
\begin{eqnarray}
 g_{t} = \begin{pmatrix}
          m_{t} & 0 & -m_{t}  \\
         -m_{t} & 0 &  m_{t}  \\
            0   &   0   & 0 
         \end{pmatrix} \,, \quad   g_{t,
   \textrm{phys}} = S_{R}^{\dagger} g_{t} S_{L} \;.
\end{eqnarray}
 
\section{Higgs boson pair production}
\label{sec:hpair}

\begin{figure}
 \centerline{
 \includegraphics{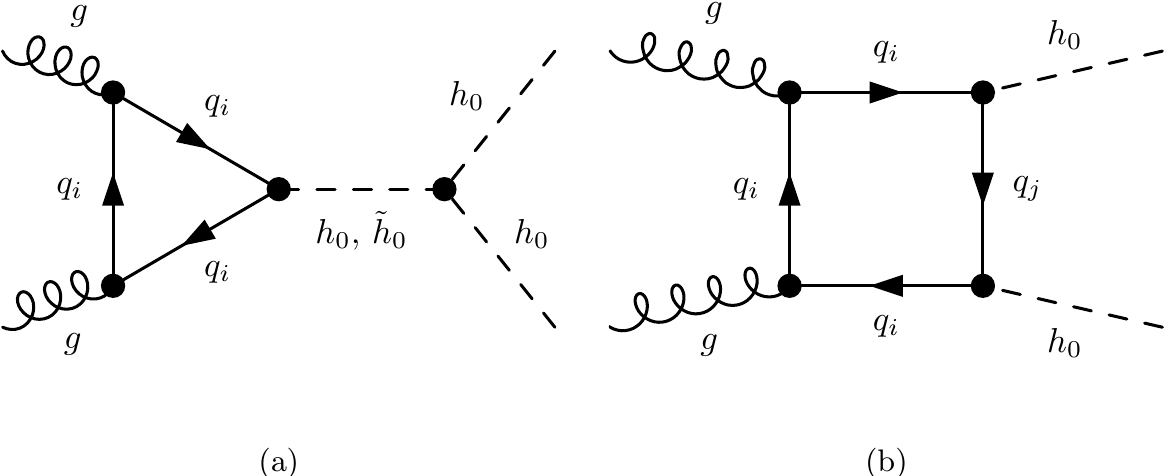}}
\caption{(a) Triangle graphs for $q= (t, \tilde{t}, \tilde{T}, b,
  \tilde{b}, \tilde{B})$ and (b) one out of six box graphs for $q_{i}, q_{j}=(t,
  \tilde{t}, \tilde{T}, b, \tilde{b}, \tilde{B})$.}
\label{fig:gghh}
\end{figure}

We shall parametrize  the $gg \to h_0 h_0$ matrix element as follows:
\begin{equation} 
\label{eq:amp}
{\cal M}(gg\to h_0 h_0) = \frac{1}{32 \pi^2}\delta^{ab} \frac{g^2}{v^2}\Big(
{\cal A}_0 P_0  + {\cal A}_2 P_2  \Big)_{\mu\nu} e(p_1)^\mu_a \, e(p_2)^\nu_b \;,
\end{equation}
with analogous conventions for $gg\to \tilde p_0 h_0$.
The pre-factor arises as follows: $1/2 \delta^{ab}$ due to  the colour trace, $1/4$ from perturbative expansion,
the fraction $g^2/v^2$ from the couplings of the vertices and $1/(4\pi^2)$
is factored out in order to give simple results for the amplitudes.
The parity-even projectors on gluon spin 0 and 2, $P_0$ and $P_2$, as well as their parity-odd counterparts, $\tilde P_0$ and $\tilde P_2$, are defined in appendix \ref{app:gghh}. 
The parton cross section for $2 \to 2$ scattering process for two massless incoming particles
is given by $1/ (16 \pi  \hat s^2) |{\cal M}|^2$  \cite{PDG}
and averaging over initial state polarizations $1/4$ and colour $1/(N_c^2-1)^2 = 1/64$ one 
arrives at\footnote{This agrees with  \cite{Glover:1987nx} with the following identifications:
$|{\cal A}_0|^2 = |\text{gauge1}|^2$ and $|{\cal A}_2|^2 = |\text{gauge2}|^2$ 
at the difference that here ${\cal A}_{0,2}$ are meant to include the LW contributions as well.}:
\begin{equation}
\frac{d\hat\sigma (gg\to h_0 h_0)}{d\hat t} =  \frac{1}{2^{19}} \frac{1}{\pi^5}\frac{g_s^4}{v^4} \,
(|{\cal A}_0|^2  + |{\cal A}_2|^2  )
\end{equation}
This result is for identical particles. In the case the particles in the final state are not 
identical one has to multiply by a factor of two\footnote{We have thus implicitly assumed that
the variable $t$ is understood to be integrated over its entire domain despite the Bose symmetry in the identical particle case.}.
The spin 0 amplitudes, parity-even and odd, receive contributions from the triangle and box diagrams, c.f. figure~\ref{fig:gghh}(left) and (right) respectively,  whereas the spin 2 amplitudes only receive contributions from the box diagrams:
\begin{equation}
{\cal A}_{0} = {\cal A}^{\triangle}_{0} + {\cal A}^{\Box}_{0} \;, \quad 
{\cal A}_2 = {\cal A}^{\Box}_2 \quad .
\end{equation}
 
For what follows it is important to notice that the gluon-quark vertex is diagonal in LW-generation space whereas the Higgs-quark vertex is not \eqref{eq:hqq}. 
Since, the Higgs-quark vertex does not contribute to the triangle graph the latter 
can be obtained from the SM with simple corrections for vertices as described 
in appendix \ref{sec:triangle}.
The modification of the box graphs are twofold. First, the external Higgs
particles are modified by the mixing factor $s_{H\!-\!\tilde H}^2$ as for the triangle. 
Second, one has to take into account that at the Higgs-quark vertex 
the LW-generations mix \eqref{eq:hqq} as discussed above. 
We find that these modifications
are most efficiently presented as follows:
\begin{eqnarray}
\label{eq:ggh0h0_box}
 {\cal A}^{\Box}_0(gg \to h_0 h_0) 
  &=&   s_{H\!-\!\tilde H}^2 
 \sum_{i,j=1}^3  \big(  f_{11}(i,j) +  f_{55}(i,j) \big)  \nonumber \\[0.1cm]
 \tilde {\cal A}^{\Box}_0(gg \to h_0 \tilde p_0) 
  &=&   - i s_{H\!-\!\tilde H} 
 \sum_{i,j=1}^3  \big(  \tilde f_{15}(i,j) +  \tilde f_{51}(i,j) \big) \;,
 \end{eqnarray}
 where 
 \begin{eqnarray}
  f_{XX}(i,j) &=& \sum_{f}^{\rm flavours} [\eta_3\,S(X)_f]_{ij}    [\eta_3 \,S(X)_t]_{ji}   (a_0)^\Box_{XX}(m_i,m_j)] 
  \nonumber  \\[0.1cm]
  \tilde f_{XY}(i,j) &=&  \sum_{f}^{\rm flavours}  [\eta_3\,S(X)_f]_{ij}    [\eta_3 \,P(Y)_f]_{ji}   (\tilde a_0)^\Box_{XY}(m_i,m_j)]  \;.
\end{eqnarray}
In regard to the formulae \eqref{eq:Aresults} it is important to notice that 
the $h_0$ and $\tilde p_0$ are associated with the 
the momenta $p_3$
and $p_4$ respectively as can be inferred from the formula in appendix \ref{sec:box}.
The couplings $S(P)_{X,Y}$, which follow from   eq.~\eqref{eq:hqq}, are:  
\begin{alignat}{2}
\label{eq:g1g5}
&S(1)_t =  \frac{1}{2}(g_t^\dagger + g_t) \;, \quad & &  S(5)_t = 
\frac{1}{2}(g_t^\dagger - g_t) \;, \nonumber \\[0.1cm]
&P(1)_t =  \frac{1}{2}(-g_t^\dagger + g_t) \;, \quad & &  P(5)_t = 
\frac{1}{2}(-g_t^\dagger - g_t) \; ,
\end{alignat}
where the top flavour was chosen as a representative and the subscript $\text{phys}$ 
has been omitted on the Yukawa couplings for the sake 
of notational brievety. 
The $\eta_3 = {\rm diag} (1,-1,-1)$ matrices 
take care of the signs of the SM and LW propagators respectively and
the couplings $g_{X,Y}$ govern the LW-generation transitions. 
The spin 2 structures  ${\cal A}^{\Box}_2$ and   $\tilde {\cal A}^{\Box}_2$ are completely analoguous.

\section{Top pair production}
\label{sec:toppair}
In this section we discuss the interference between the QCD background and resonant particles in top pair production in a qualitative manner.\footnote{We note that in ref.~\cite{Barcelo:2010bm} the authors explored these types of interferences 
in the context of minimal supersymmetric standard model and Little Higgs models.}
In the LWSM potential resonant particles, that couple to the top triangle loop 
and decay into top pairs are the $h_0$, $\tilde h_0$, $\tilde p_0$, $Z$ and $\tilde Z$ 
corresponding to the diagrams shown in figure~\ref{fig:gghh}(a) and figure~\ref{fig:gghp}(a,b,c),
respectively, with the Higgs final states replaced by top pairs. 
Here we shall neglect the $Z$ and the $\tilde Z$ as the former 
is far off-shell at $s > 2 m_t$ and the latter is severely constrained by di-lepton searches 
to be heavier than  $1\;{\rm TeV}$ and by electroweak precision measurement to be in the 
multi-TeV range. 
The corresponding amplitudes, which consist of triangle graphs only,
are easily obtained from the one for Higgs-production and are given in appendix
\ref{app:ttresults}.
 
\begin{figure}
\centerline{ \includegraphics[scale=0.7]{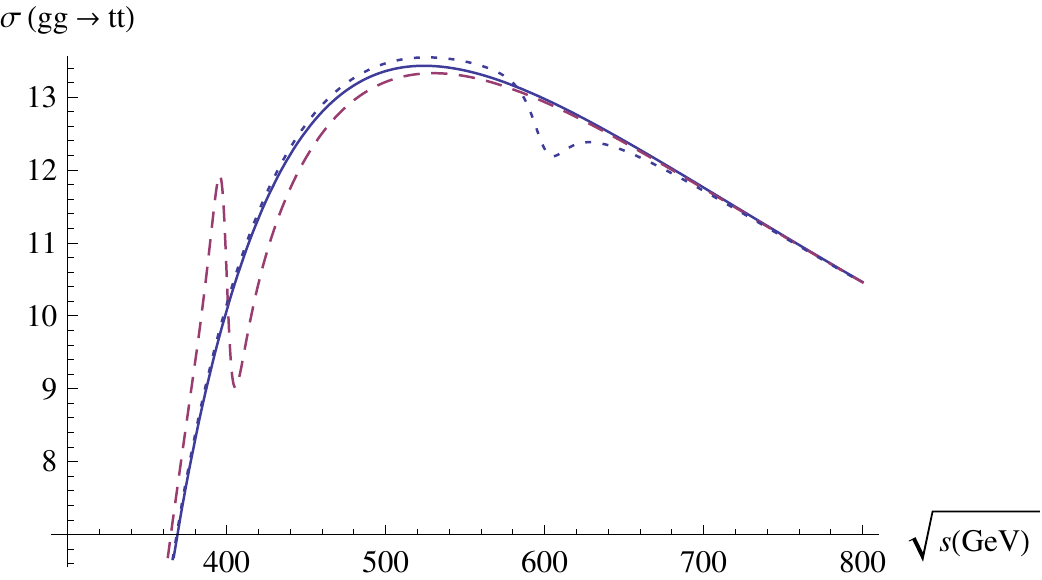}
  \includegraphics[scale=0.7]{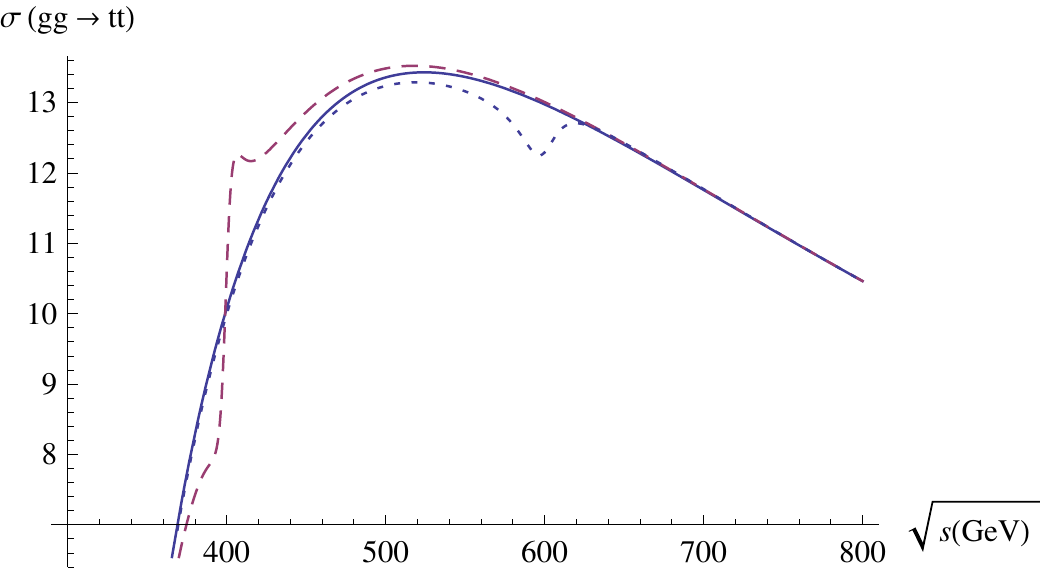}}
 \caption{The cross section $\sigma(gg \to \bar tt)$ as taken from \cite{Dicus:1994bm} qft
with energy dependent width. The solid line is the LO QCD contribution. The dashed(dotted) lines correspond to a resonance mass $m_R = 400(600) \;,{\rm GeV}$.
The left(right) figure corresponds to the usual (LW) resonance-type.}
\label{fig:peak-dip-peak}
\end{figure}

The interference effect of an intermediate resonance $gg \to R \to \bar tt$, where 
$R$ stands for a generic resonance,  
takes the following form \cite{Dicus:1994bm}:
\begin{eqnarray}
\frac{d \hat{\sigma}}{ds}(gg \to \bar tt)|_{\rm interference} &=& -|c(s)| {\rm Re}
\left[ \frac{l_\triangle}{s- m_R^2 + i m_R \Gamma_R} \right]  
 \nonumber \\[0.1cm]
&=& -|\tilde{c}(s)|\left(  (s -m_R^2)   {\rm Re}[l_\triangle] + m_R \Gamma_R  {\rm Im}[l_\triangle] \right) \;,
\end{eqnarray}
where $l_\triangle=l_\triangle(s/4 m_t^2 )$ is the appropriate triangle loop function, $c(s)$ is a well-known function of $s$ \cite{Dicus:1994bm}, $\tilde{c}(s)$ differs from $c(s)$ by a constant and $s$ the  invariant mass of the two gluons  entering the loop. If there is no loop function then the term above will lead to
a  peak-dip structure passing from constructive to destructive interference at $s = m_R^2$.
The loop-function does not change this pattern in the case where the 
resonance is a scalar or a pseudoscalar \cite{Dicus:1994bm} as the real and 
imaginary part of the loop function are positive.
The pattern persists for a 
spin-$1$ particles as well as can be inferred from the plots in 
reference  \cite{Frederix:2007gi}. Thus the question what happens in the LW case.
Due to the negative sign of the propagator and the width,
\begin{eqnarray}
\label{eq:LWttinter}
\frac{d \hat{\sigma}}{ds}(gg \to \bar tt)|_{\rm LW-interference} &=& -|c(s)| {\rm Re}
\left[ \frac{-l_\triangle(s/4 m_t^2 )}{(s- m_R^2) - i m_R \Gamma_R} \right]  \nonumber \\[0.1cm]
&=& -|\tilde{c}(s)|\left(  -(s -m_R^2)   {\rm Re}[l_\triangle] + m_R \Gamma_R  {\rm Im}[l_\triangle] \right) \;,
\end{eqnarray}
the $(s-m_R^2) {\rm Re}[l_\triangle]$  term flips sign\footnote{It is crucial that the intermediate resonance couples to the tops from the loops and the final state tops as 
otherwise a minus could be absorbed in either one of the couplings.}. 
Assuming that neither the width nor the imaginary part of the loop function 
$l_\triangle$ are anomalously large,
this  leads to a dip-peak structure.
In fact the passage from destructive to constructive interference,
which we shall call ${\cal M}_R$,  
does not coincide with the exact location of the resonance:
\begin{equation}
  {\cal M}_R^2 
 = m_R^2 + \frac{{\rm Im}[l_\triangle]}{{\rm Re}[l_\triangle]} m_R \Gamma_R
\end{equation}
Examples of the effect are shown in figure~\ref{fig:peak-dip-peak}.
The dip-peak structure is a unique feature of LW field theories, produced via gluon fusion through the top triangle,  in the case of a well isolated resonance. 
We would like to add to that in the case where the masses of two resonances are close to  each other their mixing  has to be taken into account by the so-called K-matrix formalism e.g. \cite{Peters:2004qw}. 
It is important to realize that a dip-peak structure is present in the $\pi$-$\pi$-scattering spectrum 
for  the $f_0(980)$ meson due to the extremely broad $f_0(600)$ ($\sigma$-meson)   \cite{Harada:1995dc}. Thus strongly coupled extensions of the SM, such as technicolor, 
might have similar signals as the LWSM. 

\subsection{A comment on the top forward backward asymmetry}
\label{sec:tFBA}
Currently, the top forward-backward asymmetry (tAFB), $A_{\rm FB}^{\bar t t} = 0.475(114)$ for $M_{tt} > 450 \,{\rm GeV}$ at \cite{Aaltonen:2011kc} at $5.3\,{\rm fb}^{-1}$, deviates from SM prediction $A_{\rm FB}^{\bar t t} = 0.088(13)$ \cite{Aaltonen:2011kc} at about the $3\sigma$-level at the TeVatron\footnote{The very recent D0-results at $5.4\,{\rm fb}^{-1}$ is much closer 
to the SM value \cite{Collaboration:2011rq}.}. The SM prediction originates from  a charge asymmetry
which, due to the fact that the TeVatron is a $p\bar p$-collider, translates into
a forward-backward asymmetry. Thus the question is whether the LWSM has the potential
to explain this discrepancy. A nice summary of perturbative approaches to the tAFB
is given in reference \cite{Cao:2010zb}. The LWSM qualifies at the same level as
a $Z'$-model with SM-like couplings, where  the role of the $Z'$ is taken by $\tilde Z$.\footnote{The LWSM does not qualify as an axi-gluon, nor are there large flavour changing couplings between the first and third generation in its minimal version.} 
We roughly get $A_{\rm FB}^{\bar tt} \simeq 0.01$, for a $m_{\tilde Z} = 1 \,{\rm TeV}$,
at best which is in the right direction but too small
to join into the current excitement.\footnote{Note $m_{\tilde Z} = 1 \,{\rm TeV}$ is even a bit low
in regard to electroweak precision data \cite{Underwood:2008cr}.}
Note, as only the absolute value
of the propagator enters, the wrong-signs of the propagator and the width do not matter.
We have not evaluated the interference of
the $\tilde Z$ and the SM $Z$, but expect it to be of similar size.

\section{Numerical results}
\label{sec:numerical}
We compute cross sections for $pp \to h_0 h_0 / \tilde p_0 h_0$ 
and the differential cross section $pp \to \bar tt$ via gluon fusion
at the LHC for $\sqrt{S}=7/14~{\rm TeV}$ respectively\footnote{The cross section for vector boson fusion $qq \to h_0 h_0 jj$ is about $2\%$ of $gg \to h_0 h_0$ and thus negligible.}. We denote the $pp$ center of mass energy by capital $S$ and the partonic center of mass energy by lower case $s$ throughout this paper. 
The renormalization and factorization scale has been chosen to be $\mu_{r}=\mu_{f}=2 m_{h_0}$ for $pp \to h_0 h_0$
and $\mu_{r}=\mu_{f}=m_{h_0}+m_{\tilde{p}_0}$ for $pp \to \tilde{p}_0 h_0$.
We use the MSTW $2008$ LO ($90\%$ C.L.) for parton
distribution functions with the strong coupling calculated to one-loop
order for $\alpha_{s}(m_{Z})=0.13939$ ~\cite{Martin:2009iq}.
We use LO predictions for $gg \to h_0 h_0/\tilde p_0 h_0$. The NLO corrections
in the later case are rather large \cite{Dawson:1998py}; almost $100\%$ as can be inferred
from figure~6 of that reference. Fortunately, the shape of the corrections are almost identical 
to the LO result and thus should not distort the analyses too much. For $gg \to \bar tt$ we also use LO predictions
with the factorization scale set to $\mu_{f}=m_{t}$ and the renormalization scale set to $\mu_{r}=m_{t}(m_{\phi})$ for $gq\bar{q}$ ($gg \phi$ for 
$\phi=\tilde{p}_{0},\tilde{h}_{0},h_0$)
couplings for which we comment in section \ref{sec:toppair}.

For the numerical computations we have used various computer packages to be referred to below.
The FeynArts~\cite{Hahn:2000kx} model file has been generated automatically using LanHEP~\cite{Semenov:2002jw}.
The resulting model files were modified to allow for wrong-sign propagators in the auxiliary field formalism.
Fortran code for the cross sections was generated with the use of
FormCalc~\cite{Hahn:1998yk}. All loop integrals were computed using LoopTools~\cite{Hahn:1998yk}.

The width-mass ratios, widths and branching ratios  for  $h_0$ and $\tilde h_0$ are depicted in 
appendix \ref{app:plots}.  in figures~\ref{fig:wcont},\ref{fig:width}(left) and \ref{fig:branch}
respectively. They will be referred throughout and serve to understand the results qualitatively.
Possibly the most important aspect for further understanding 
is that the width of the $\tilde h_0$ (in figure~\ref{fig:width}(left) appendix \ref{app:plots}) raises significantly when the $t \bar t$-threshold is crossed (in parameter-space $m_{\tilde h_0} > 2 m_t$) and is relevant for the triangle diagrams with intermediate $\tilde h_0$.

\subsection{Contraints on LW mass scales}
\label{sec:scales}
Before presenting the main results 
the new  LW scales have to be discussed. 
There are six LW mass scales plus the mass of the SM-like Higgs boson out of which 
five are constrained to be rather high and generally do not impact on our investigation
 The parameters are:
\begin{itemize}
\item The scales $M_1$ and $M_2$ of the LW gauge bosons associated 
with U(1)$_Y$ and SU(2)$_L$ are constrained by electroweak precision measurements 
to be in the multi-TeV range \cite{Underwood:2008cr}. 
We shall set $M_1 = M_2 = 1\,{\rm TeV}$ throughout this paper  as in this range the masses have no major influence on our results.
\item
The fermion mass scales $M_Q$, $M_u$ and $M_d$   \footnote{Due to 
chiral suppression, only heavy flavours are relevant. This statement can be inferred from 
the HD formalism. Thus only the top and the beauty quarks are taken into consideration.} 
are constrained through loop-contributions to electroweak precision measurements 
 to be in the multi-TeV range \cite{Chivukula:2010nw}. 
 For the $gg \to h_0 h_0$ channel the fermion mass scale has little influence for 
 $M_Q= M_u= M_d > 500 \,{\rm GeV}$ as can be inferred from the 
 appendix \ref{app:plots} figure~\ref{fig:mh2}.  
There are no qualitative changes when one goes away from the limit of equal masses 
and we therefore assume the the fermion mass scales to be  $500 \,{\rm GeV}$ in the plots.
For the $gg \to \tilde p_0 h_0$ channel there are some threshold effects due to 
the box diagrams.
 \item The masses of the  two neutral CP-even Higgses $h_0$ and $\tilde h_0$ \eqref{eq:Htable}. Every other parameter, in the Higgs sector, can be expressed in terms of these two. In particular 
the pseudoscalar mass, at tree-level satisfies \eqref{eq:Htable}
\begin{equation}
\label{eq:tree}
m_{\tilde p_0}^2 = m_{h_0}^2 + m_{\tilde h_0}^2 \;,
\end{equation}
where we have dropped the subscript ``phys'' and shall do so 
in the remainder of this paper.
The Higgs parameter-space has already been studied in other works.
The collider analysis of $gg \to h_0 \to \gamma \gamma$  \cite{Krauss:2007bz} 
was extended to   final state channels $\gamma Z$ and $WW$ in \cite{Alvarez:2011ah}.
A  part of the parameter-space has been found to be 
excluded by TeVatron results, c.f. figure~3 in that paper.  It has to be added that this work
was done in the narrow width approximation. Inspecting the plots in figure~\ref{fig:wcont} 
it would seem that the effect of the width should be moderate in most of the parameter-space that has been excluded.  The overlap of the interesting parameter-space and their excluded region is rather small and we leave it to the reader to convince him or herself of this
fact.
Using the correspondence of the the LWSM Higgs-sector and the type-II two Higgs doublet model mentioned, in the introduction, 
the effects of the charged Higgs boson $\tilde h_+$ on flavour physics were investigated 
in reference \cite{Carone:2009nu}.
Using NLO predictions 
for $b \to s \gamma$, neglecting the influence of all other LW states, 
which is  consistent with our analysis, 
it was found that $m_{\tilde h_+} > 463\,{\rm GeV}$ at the $95\%$ confidence level.
Together with  the tree-level relation \eqref{eq:tree} and $m_{\tilde p_0} = m_{h_+}$ \eqref{eq:Htable}
this sets a significant constraint on the lower range of our parameter space. 
Concerning this indirect bound there are two remarks to be made.
First, the individual theoretical uncertainties were added in quadrature, which is common practice, and thus the uncertainty might be considered to be a little bit on 
the low side. 
Second, the tree-level relation between the Higgs masses might receive 
significant radiative corrections due to the large top mass 
which is the case in the MSSM.
\end{itemize}

We would like to add that  the limit of degenerate masses of the $h_0$ and $\tilde h_0$, 
parametrized by $r_{h_0} \equiv m_{h_0}/m_{\tilde h_0}$, is somewhat delicate. 
In connection with real particles, in the sense of parton level, it does not make sense
to treat them separately. This can be seen in the pole in $r_{h_0}$ in 
$s_{H - \tilde H} = (1+r_{h_0}^2)(1-r^4_{h_0})^{-1/2}$.
For virtual particles it is best to resort to the HD-formalism where everything should remain 
consistent.
In regard to these points we disregard the parameter space where 
\begin{equation}
\label{eq:pert}
r_{h_0} > 0.8 \;,  \qquad r_{h_0} \equiv \frac{m_{h_0}}{m_{\tilde h_0}} \;,
\end{equation}
which is somewhat more conservative than the value $r_{h_0} > 0.9$ chosen in
\cite{Alvarez:2009af}.
It would be interesting to study these effects, from scratch, in the HD-formalism and 
find the relation to the K-matrix formalism 
\cite{Peters:2004qw} used to improve on two nearby Breit-Wigner resonances in usual field theory.

\subsection{Results for $gg \to h_0 h_0$}
The main point is that for $m_{\tilde h_0}$ slightly above  $2 m_{h_0}$ the cross
section is large, three orders of magnitude larger than the one of the SM, dominated by the resonant contribution in the triangle graph 
figure~\ref{fig:gghh}(left). This is reminiscent of the situation in the MSSM \cite{Plehn:1996wb}. 
\begin{figure}
\centerline{
 \includegraphics[scale=0.6]{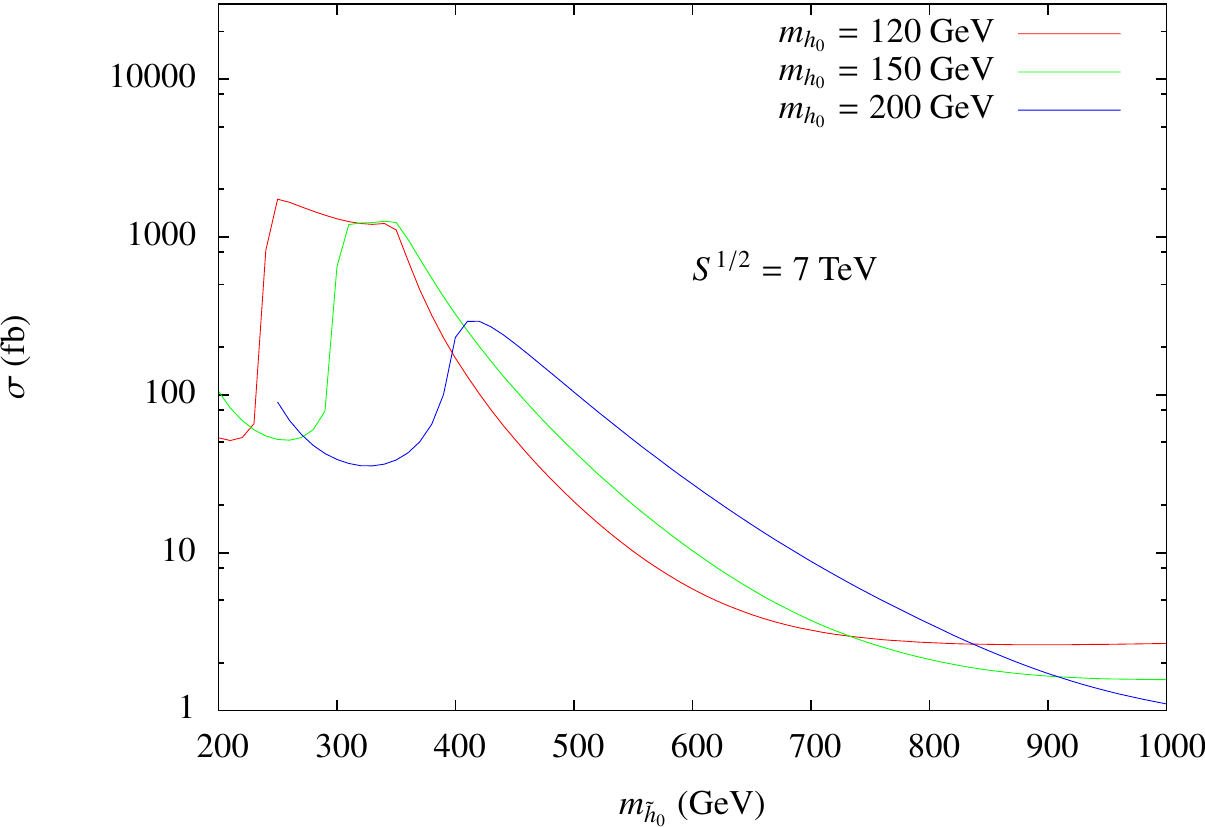} 
 \includegraphics[scale=0.6]{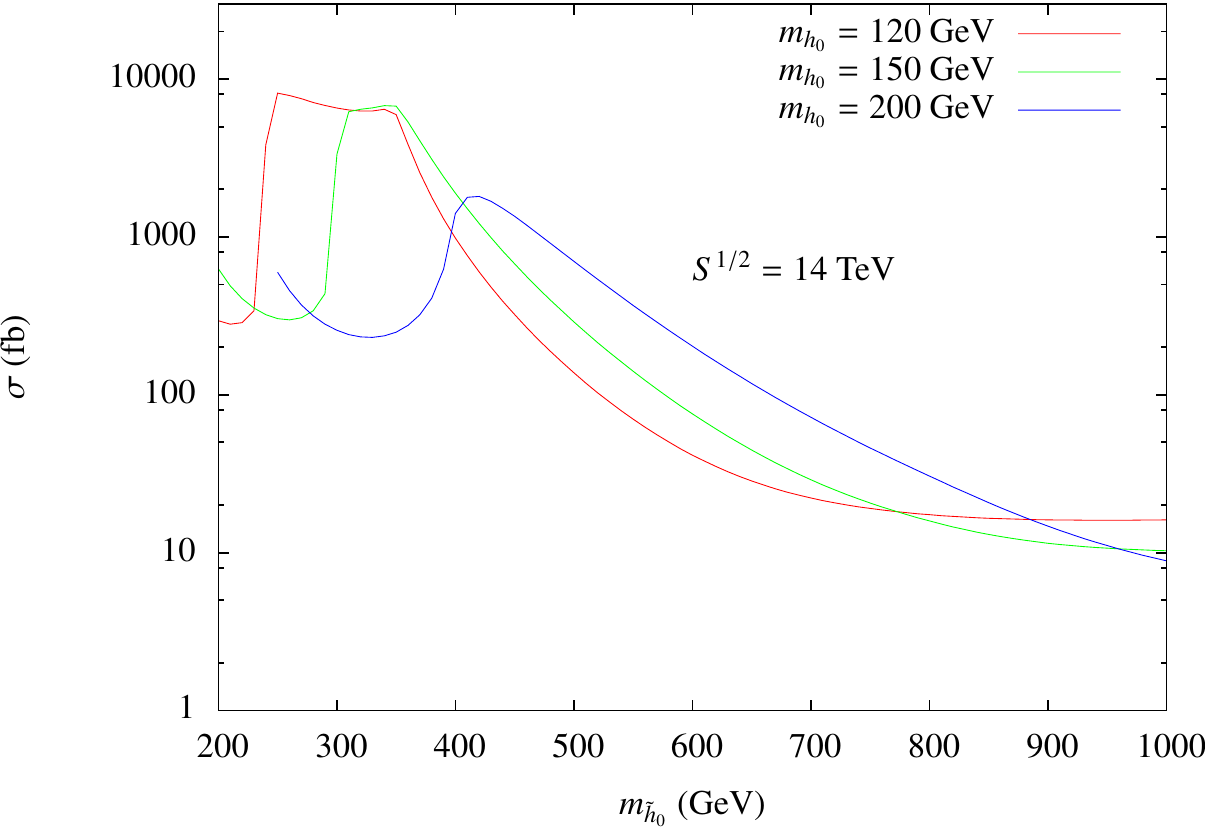}}
\caption{\small The cross section (in fb) for $gg \to h_0 h_0$ via gluon fusion at
  the LHC for $\sqrt{S}=7/14$ TeV  respectively versus 
  the mass of the $\tilde h_0$, $m_{\tilde h_0}$, for three different values of $m_{h_0}$.}
\label{fig:first}
\end{figure}

\begin{figure}
 \centerline{
 \includegraphics[scale=0.7]{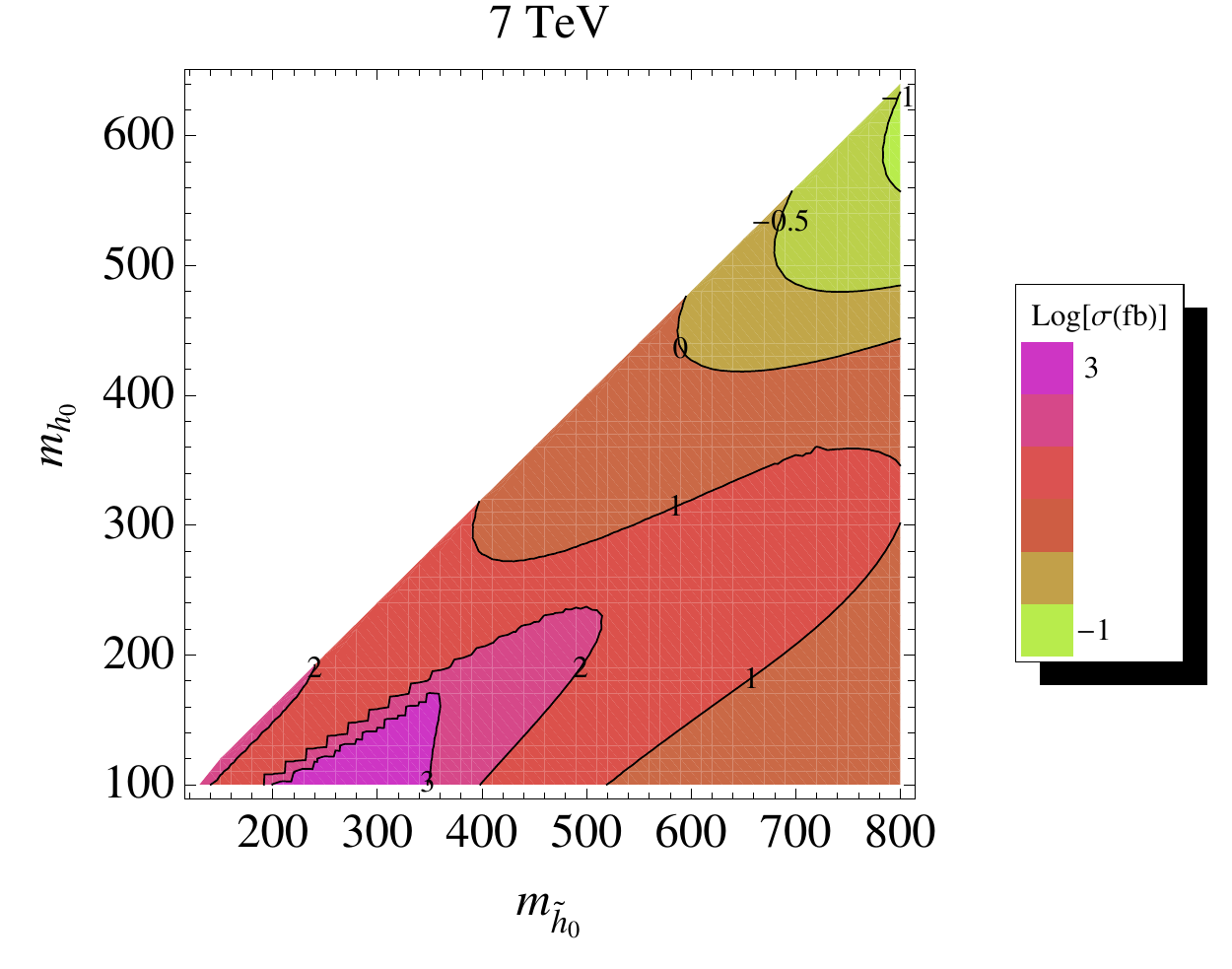}
 \includegraphics[scale=0.7]{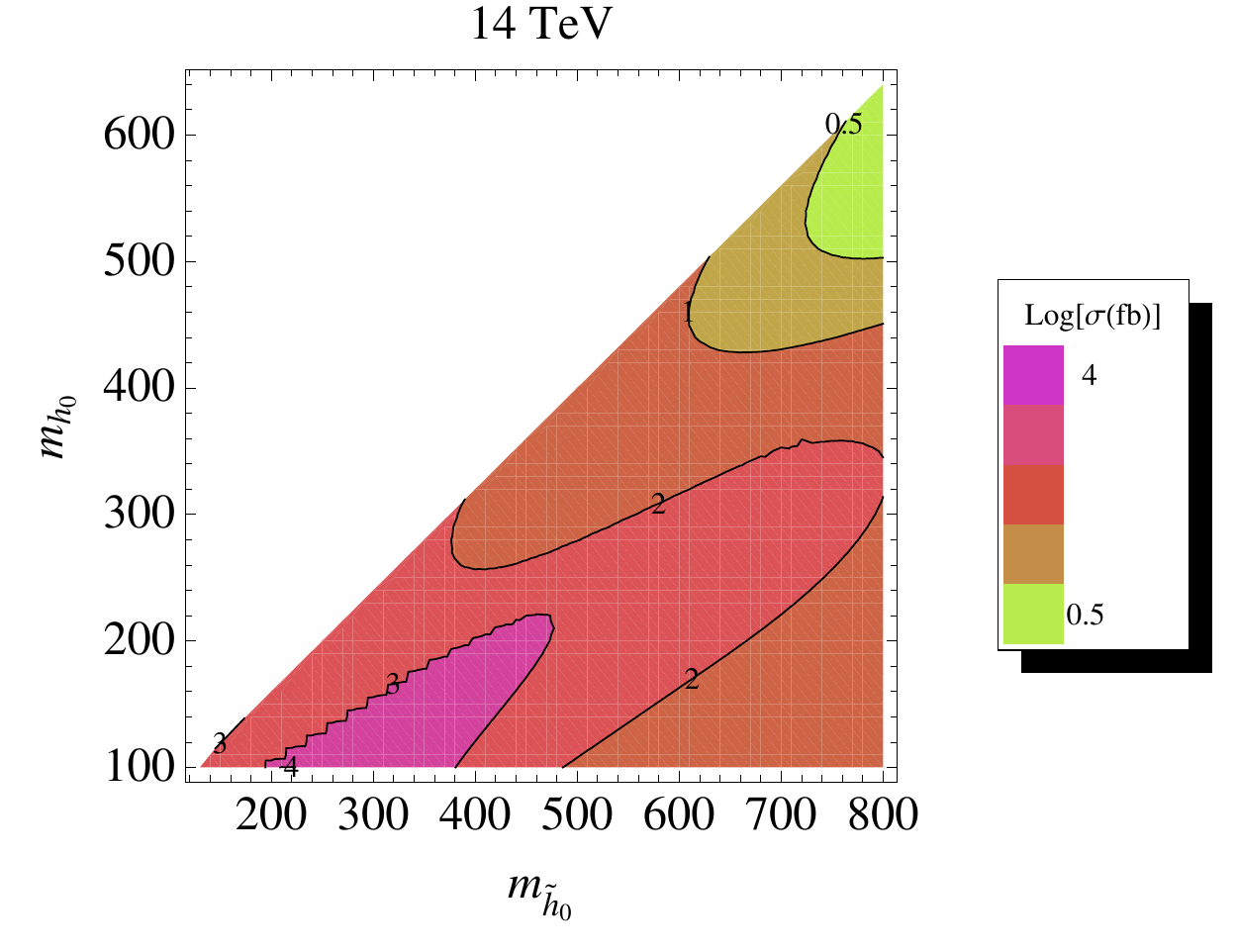}
}
\caption{Contour plot of the total cross section (if fb) for $gg \to h_0 h_0$ ($ \sqrt{S} = 7/14$ TeV respectively) versus the light Higgs boson mass, $m_{h}$ and 
heavy Higgs boson mass, $m_{\tilde{h}}$ for $M_{Q}= 0.5$ TeV and $M_{1} = M_{2} =1$ TeV. 
Note figure~\ref{fig:first} corresponds to horizontal sections in this plot.}
\label{fig:contour}
\end{figure}

In figure~\ref{fig:first} we show the total cross section for $gg \to
h_0 h_0$ for $\sqrt{S}=7/14~{\rm TeV}$ respectively
as a function of $m_{\tilde h_0}$ for three different values of 
$m_{h_0}$.  More detailed information can be 
inferred from the contour plots in the $(m_{h_0} , m_{\tilde h_0})$-plane 
shown in figure~\ref{fig:contour}.
As mentioned above one observes a sharp raise of the cross section 
when the LW Higgs mass crosses the threshold  $2 m_{h_0}$, c.f. figure~\ref{fig:first}.
For higher  $m_{\tilde h_0}$ the resonance contribution decouples and finally approaches
the SM value. An interesting effect arises when the top threshold is reached. 
For the observation to be made below recall that the process is dominated 
by the triangle graph with  an intermediate LW Higgs propagator 
of the form $(s - m_{\tilde h_0}^2 -i  m_{\tilde h_0} \Gamma_{\tilde h_0})^{-1}$.
The slight dip in the branching ratio, c.f. figure~\ref{fig:branch}(right), 
below the $t \bar t$-threshold results in a slight raise of the curve 
in case the where  $m_{h_0} < m_t$.  Once the $t \bar t$-threshold is reached the rapidly 
growing decay rate is damped through the additional relevant part in 
the propagator. Note the lower part of the blue curve raises.
In the HD-formalism  this can be understood by the to the two poles $m_{h_0}$ and $m_{\tilde h_0}$ approaching each other.

\subsection{Results for $gg \to h_0 \tilde p_0$}

\begin{figure}
\centerline{
 \includegraphics[scale=0.6]{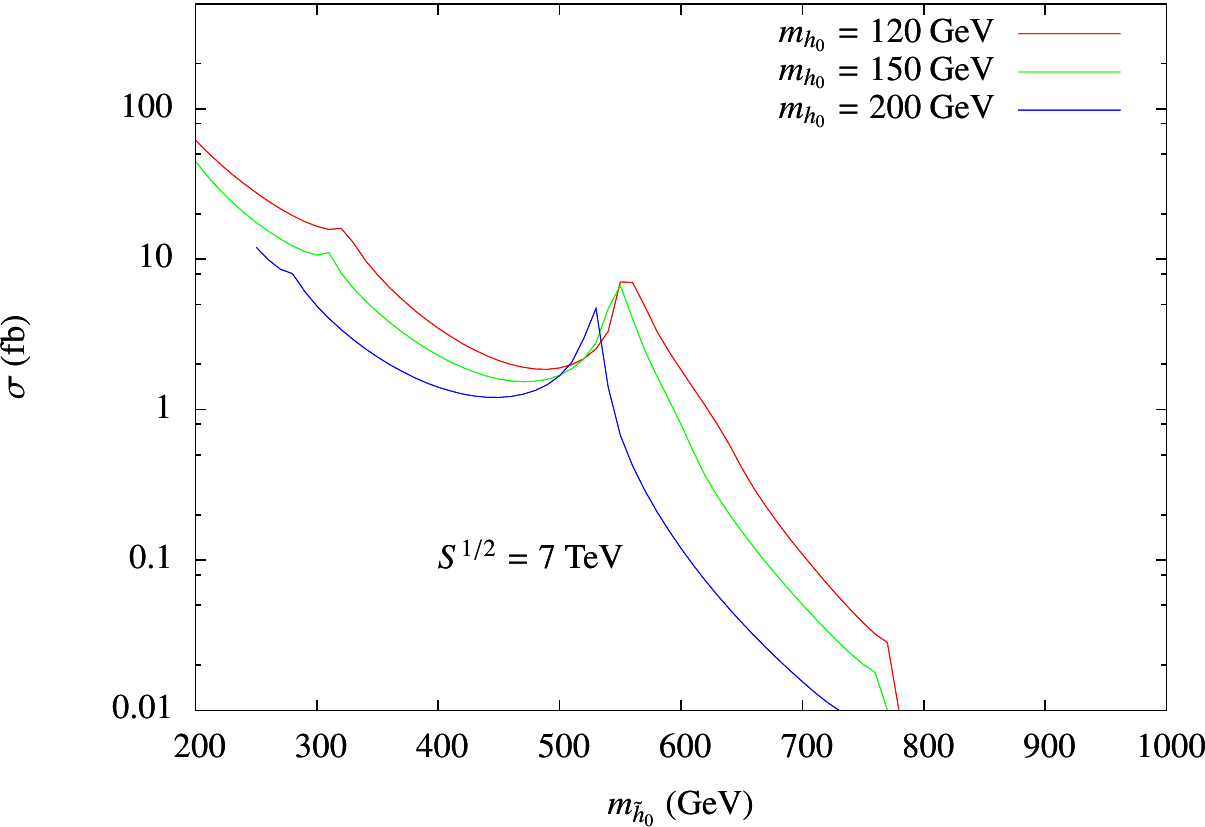} 
 \includegraphics[scale=0.6]{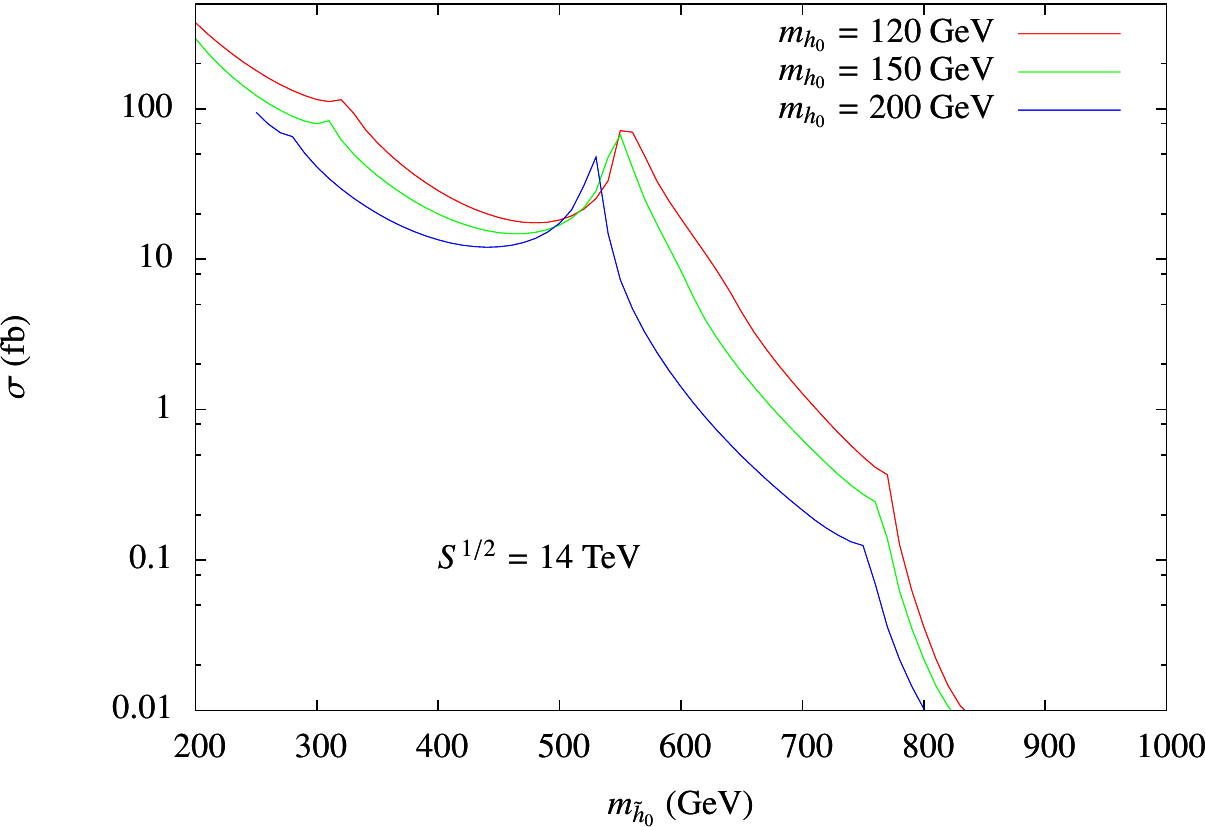}}
\caption{\small The cross section (in fb) for $pp \to h_0 \tilde p_0$ via gluon fusion at
  the LHC for $\sqrt{S}=7/14$ TeV respectively versus 
  the mass of the $\tilde h_0$, $m_{\tilde h_0}$, for three different values of $m_{h_0}$.
  Note the kinks are due to crossing thesholds in corners of the box diagrams as described in the text; recall 
  $M_Q = M_u = M_d = 500~{\rm GeV}$.}
\label{fig:ggph}
\end{figure}

\begin{figure}
 \centerline{
 \includegraphics[scale=0.7]{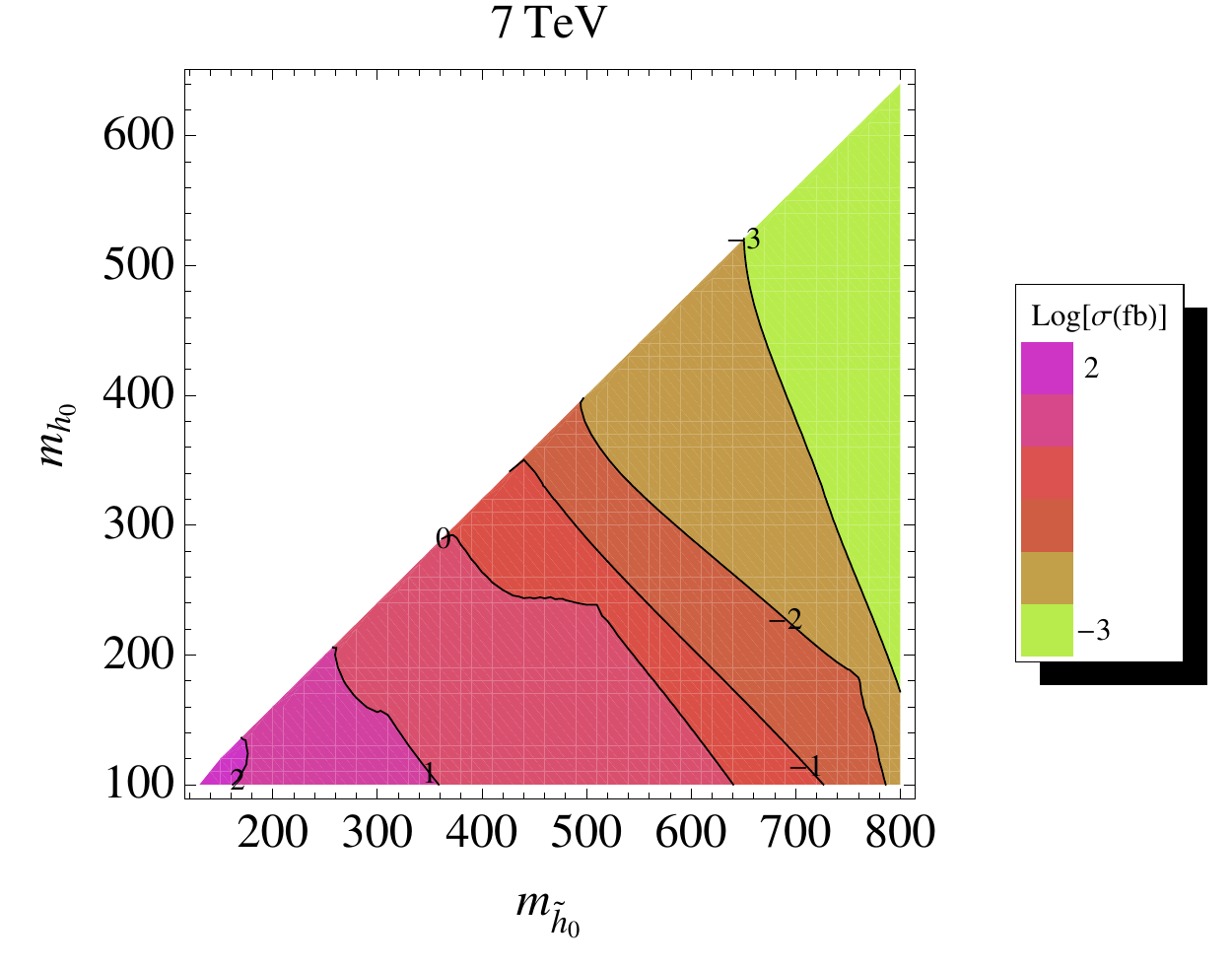}
 \includegraphics[scale=0.7]{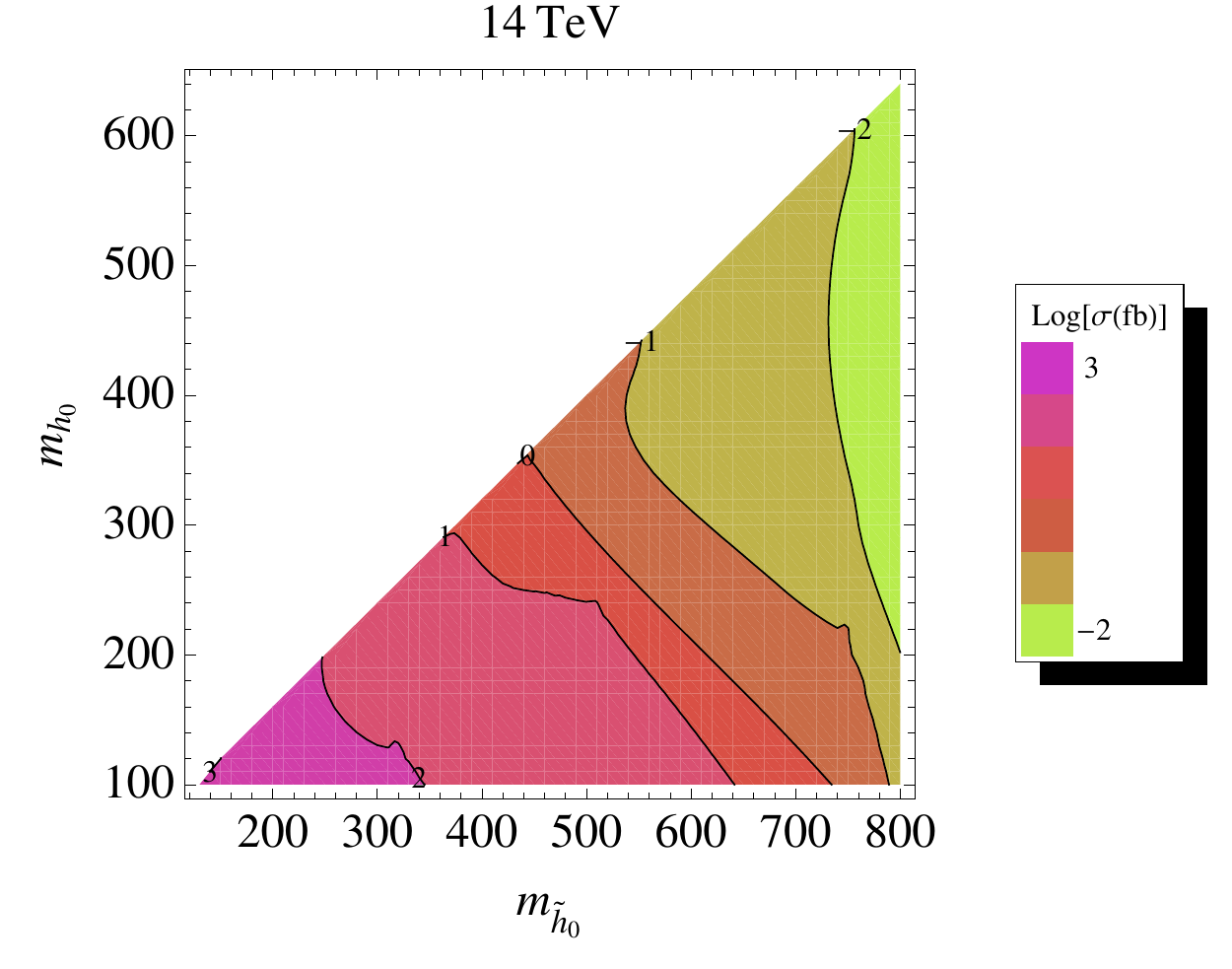}
}
\caption{Contour plot of the total cross section (in fb) for $gg \to h_{0} \tilde p_{0}$ ($ \sqrt{S} = 7/14$ TeV respectively) versus the light Higgs boson mass, $m_{h}$ and 
heavy Higgs boson mass, $m_{\tilde{h}}$ for $M_{Q}= 0.5$ TeV and $M_{1} = M_{2} =1$ TeV. 
}
\label{fig:contour1}
\end{figure}

The cross section for $\sqrt{S} = 7/14 \,{\rm TeV}$ respectively with fixed $m_{h_0}$ are shown in figure~\ref{fig:ggph}.
The corresponding contour plots are shown in figure~\ref{fig:contour1}.
The crucial difference to $gg \to \tilde h_0 \to h_0 h_0$, in terms of the triangle diagram,  is that there's no parameter
region where there's a dominant resonance effect.
The   diagrams are shown in figure~\ref{fig:gghp}: the intermediate $Z$ and $\tilde Z$ 
are either too light or too heavy respectively and the process $gg \to \tilde p_0 
\to \tilde p_0 h_0$ is not on-shell. There's a remnant of the latter
effect when the $m_{h_0}$ is relatively small and $p_0 
\to \tilde p_0 h_0$ approaches an on-shell configuration.
The cross section is enhanced for  
$r_{h_0} \to 0.8$ \eqref{eq:pert} due a larger coupling $s_{H-\tilde H}$ of the SM-like Higgs
to the two pseudoscalars. For large $m_{\tilde h_0}$ the cross section
goes to zero which is consistent with the fact that this process is not present in the SM.
We further note the thresholds in $2 m_t$ and $ m_t + m_{\tilde t}$ 
in the pseudoscalar mass, parametrized in 
terms of the CP-even Higgs masses according to eq.~\eqref{eq:tree}, become visible. These effects are not present in $gg \to h_0 h_0$, since, the mass of the final state particles was assumed to be below these thresholds.

\subsection{Results for $gg \to \bar t t$  (the $M_{tt}$-spectrum)}

\begin{figure}
 \centerline{
 \includegraphics[scale=0.7]{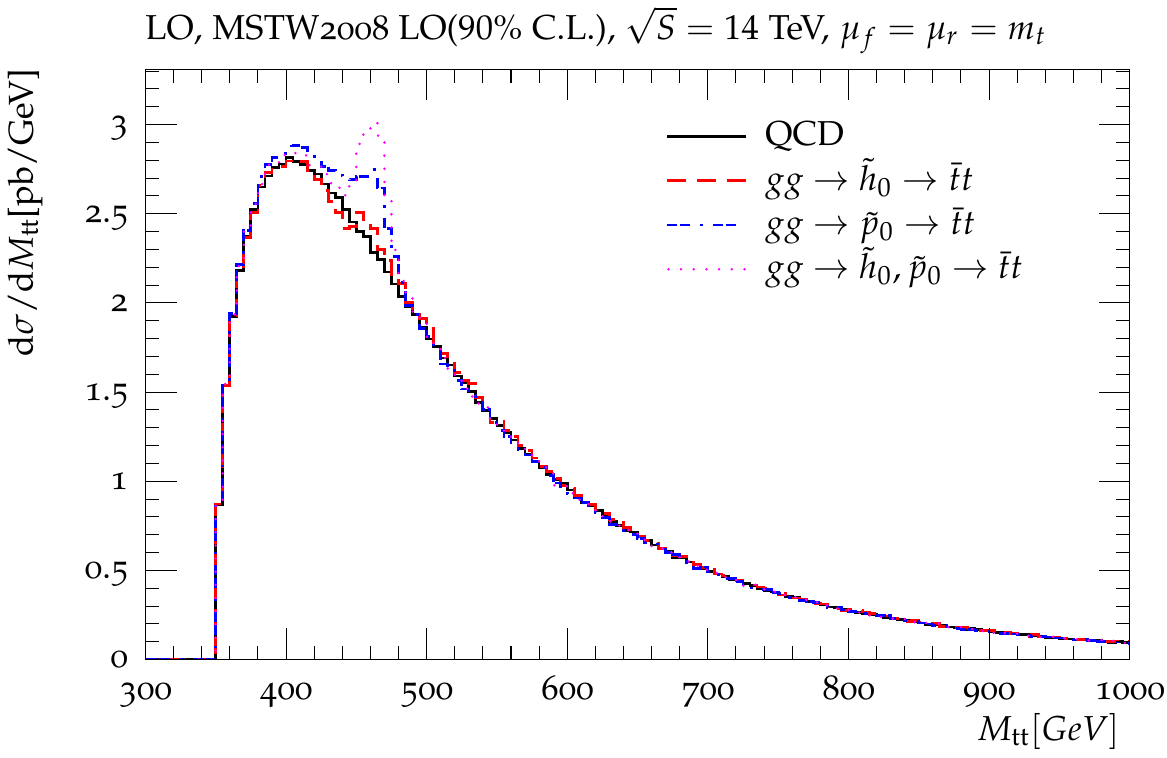}
 \includegraphics[scale=0.7]{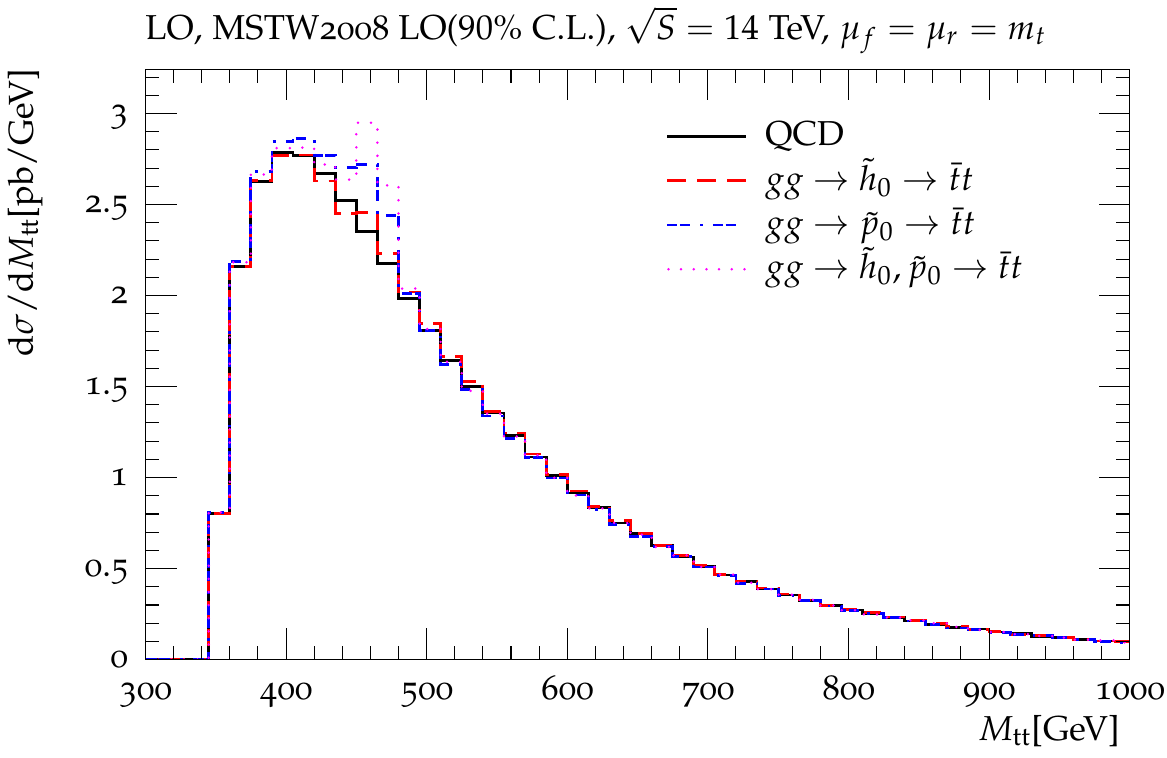} 
}
\centerline{
\includegraphics[scale=0.7]{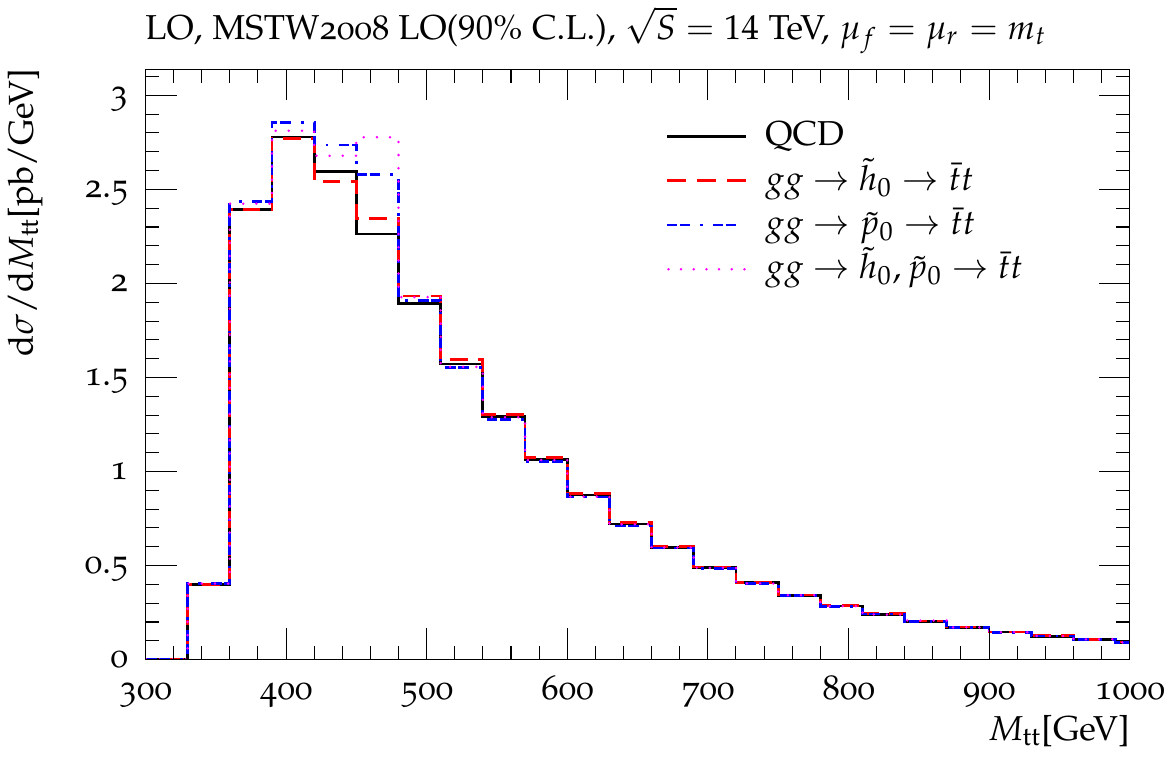}
\includegraphics[scale=0.7]{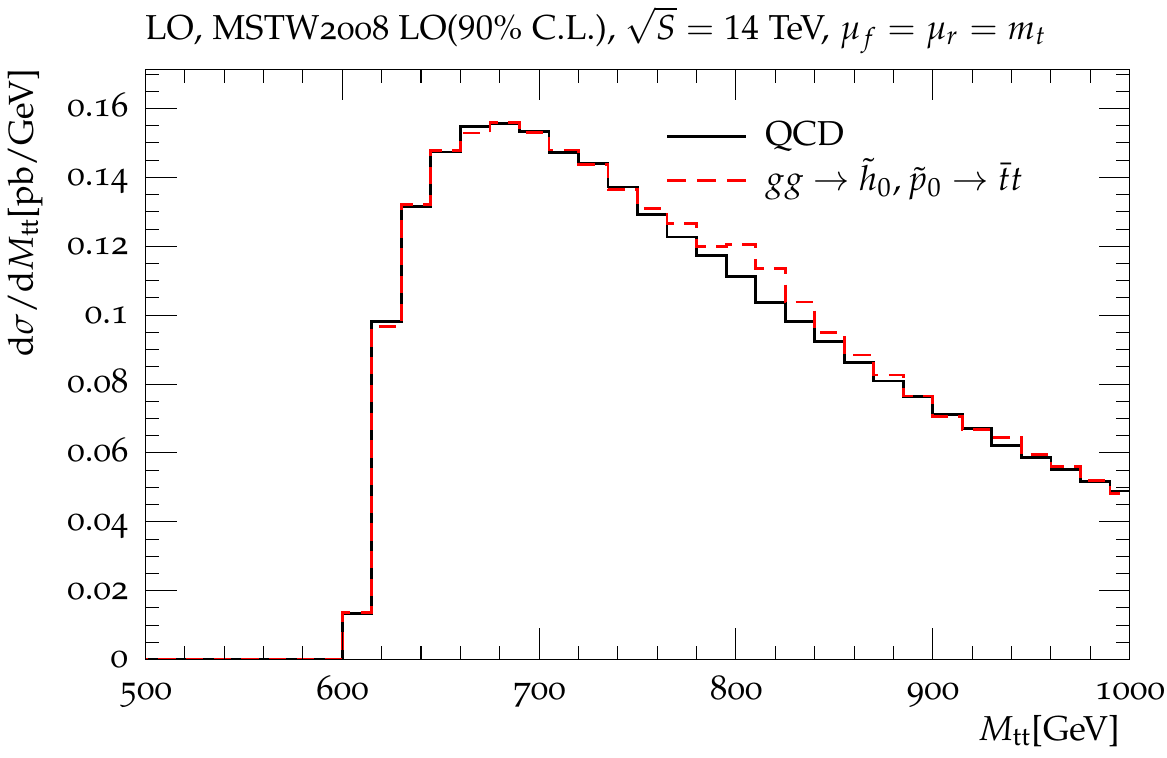}
 }
\caption{Histograms of the top pair invariant mass, $M_{\bar tt}$, for $gg \to \bar tt$ for $\sqrt{S} = 14\,{\rm TeV}$. 
 (top left)(top right) and (bottom left)  for $5/15/30\,{\rm GeV}$-bins, respectively.
 A dip-peak structure is to observed by 
the interference of the QCD-background with LW Higgs states. 
In these figures we have chosen the following mass values $m_{\tilde h_0} = 450\;{\rm GeV}$ 
and $m_{\tilde p_0} = 467 \;{\rm GeV}$ which implies with 
eq.~\eqref{eq:tree} $m_{h_0}= 125\;{\rm GeV}$. (bottom right) 
We plot $M_{tt}$ for $m_{\tilde{h}_0}=800\;{\rm GeV}$ in $15~{\rm GeV}$ bins where we assume $M_Q = M_u = M_d = 500\; {\rm GeV}$ where 
the signal to background ratio is significantly improved by  
$p_T$-cut of $250\,{\rm GeV}$  to each top.}
\label{fig:mtt_histogram}
\end{figure}

In this section we present the $\bar t t$-mass spectrum. 
In the case where the $h_0$ or $\tilde p_0$ are above 
the $\bar tt$-threshold ($m_{h_0},m_{\tilde p_0} > 2 m_t$) 
a dip-peak structure is to be expected, originating from 
the interference of the QCD-background with LW Higgs states, 
as described in section \ref{sec:toppair}. This phenomenon is observed in the actual simulation 
as can be inferred from  figure~\ref{fig:mtt_histogram} for $m_{h_0},m_{\tilde h_0} =  (125,450) \,{\rm GeV}$ but is hard to see for higher values of LW Higgs mass e.g.
$m_{h_0},m_{\tilde h_0} =  (125,800) \,{\rm GeV}$. This is because the width of the intermediate $\tilde h_0$ and $\tilde p_0$ 
becomes large and the two terms in eq.\eqref{eq:LWttinter} tend to cancel each other. In the latter case the signal to background ratio can be improved significantly in the case where a $p_T$-cut of $250\,{\rm GeV}$ is applied to each top c.f. figure~\ref{fig:mtt_histogram}. This study could be explored further using the top tagger of ref.~\cite{Plehn:2009rk}, since, the transverse momentum of the top quarks peak around $300$ GeV, i.e., the tops are boosted\footnote{The search strategies outlined in refs.~\cite{Baur:2008uv,Baur:2007ck,Barger:2006hm} can, also, be applied here as well.}.  For $\tilde{h}_0$ masses in the multi-TeV range one could employ the top tagging methods of ref.~\cite{Kaplan:2008ie}.\footnote{For a review of top tagging we refer the reader to ref.~\cite{Abdesselam:2010pt}} 
Note that the two LW-states $\tilde h_0$ and $\tilde p_0$ are necessarily close to each other 
in case of a low SM-like Higgs mass 
by virtue of the tree-level relation \eqref{eq:tree}. The effect of which can be seen in figure~\ref{fig:mtt_histogram} where the individual  parts are given.
We have chosen $M_{\bar tt}$-bins of $5$, $15$ and $30\;{\rm GeV}$ respectively 
 for $\sqrt{S} = 14\;{\rm TeV}$.
A bin-size of $5\;{\rm GeV}$ seems unrealistic (in view of detector resolutions), whereas
$15\;{\rm GeV}$ can be achieved and $30\;{\rm GeV}$ might very well be
the reference value for early publications. 
A fundamental particle is described by its mass, spin and to some extent its 
interactions. So far we have not addressed the spin. 
The latter can be determined, as usual, through  angular distributions. 
In \cite{Frederix:2007gi}, c.f. figure 15, the so-called Collins-Soper angle is advocated as promising observable. 

We would like to add that the simulations were performed with LO order QCD backgrounds. 
For an assessment of NLO corrections we refer the reader to figure~2 in \cite{Frederix:2007gi}.
Besides the fact that they are not too large in the low mass region the important thing is that
the shape is very similar to LO and thus very different to a resonant structure. 
In regard to the values of the $d\sigma(gg \to \bar tt)/dM_{\bar tt}$ differential cross section 
it should be  kept in mind that it is not the top-pair that is observed 
in the detector. The efficiency of the top-reconstruction is estimated to be about $5\%$ \cite{Aad:2009wy,Cogneras:974640}. 
The effects of the Higgs resonances for $\sqrt{S} = 7\;{\rm TeV}$ seem to small to be observed 
and we have relegated the corresponding plot to appendix \ref{app:plots} figure~\ref{fig:width}(right).  In that case the gluon density is too small and 
 $\bar qq \to g \to \bar tt$ becomes more important. The latter being in a color octet 
 representation, does not interfere with the LW
 contributions which is in a color singlet representation which leads to a reduction of the effect.  

\section{The $gg \to h_{0} h_{0} \to b \bar{b} \gamma \gamma$ channel at the LHC}
\label{sec:signal}

In this section we will access the observability for double Higgs boson production in the LWSM \footnote{The $h_{0} h_{0} \to \gamma \gamma b \bar{b}$  channel has been studied in the past in the context of the Randal--Sundrum model by both ATLAS~\cite{Azuelos:2002fv} and CMS~\cite{Gennai:2004jq,Dominici:2004ti}, in the SM and MSSM~\cite{RichterWas:1996ak, Baur:2003gp}, and, most recently, in the context of a hidden sector Higgs boson~\cite{Bowen:2007ia}. 
} being the more promising than the $\tilde{p_0} h_0$-channel, for light Higgs boson masses in the range of $\sim 120 - 130~{\rm GeV}$ in the $gg \to h_0 h_0 \to \gamma \gamma b \bar{b}$ channel. This channel is of particular relevance, since, searches for the SM Higgs boson at ATLAS
exclude SM Higgs boson masses at $95\%$ C.L. in the range $155 -190$ GeV and $295-450$ GeV \cite{AtlasHiggs} and at CMS exclude SM Higgs boson masses at $90\%$ C.L. in the range $145 - 480$ GeV \cite{CMSHiggs} \footnote{These bounds apply to the SM. For the LWSM we would expect, from the viewpoint of the HD-formalism, very similar or slightly stronger bounds.}.  This suggests the SM-like Higgs boson should reside in the low mass region, i.e, $m_{h_0} < 145~{\rm GeV}$.

\begin{figure}
\centerline{
 \includegraphics[scale=0.7]{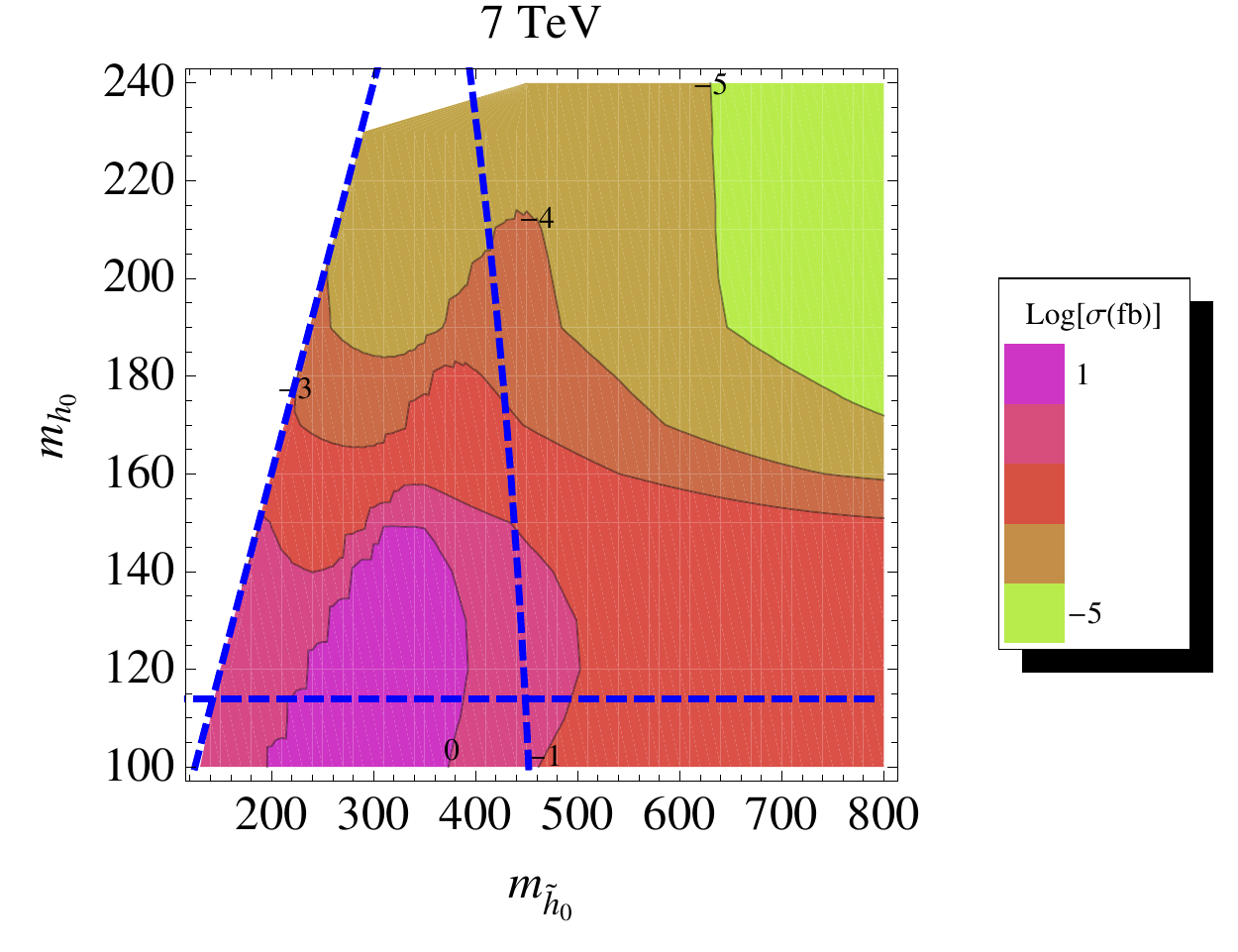}
  \includegraphics[scale=0.7]{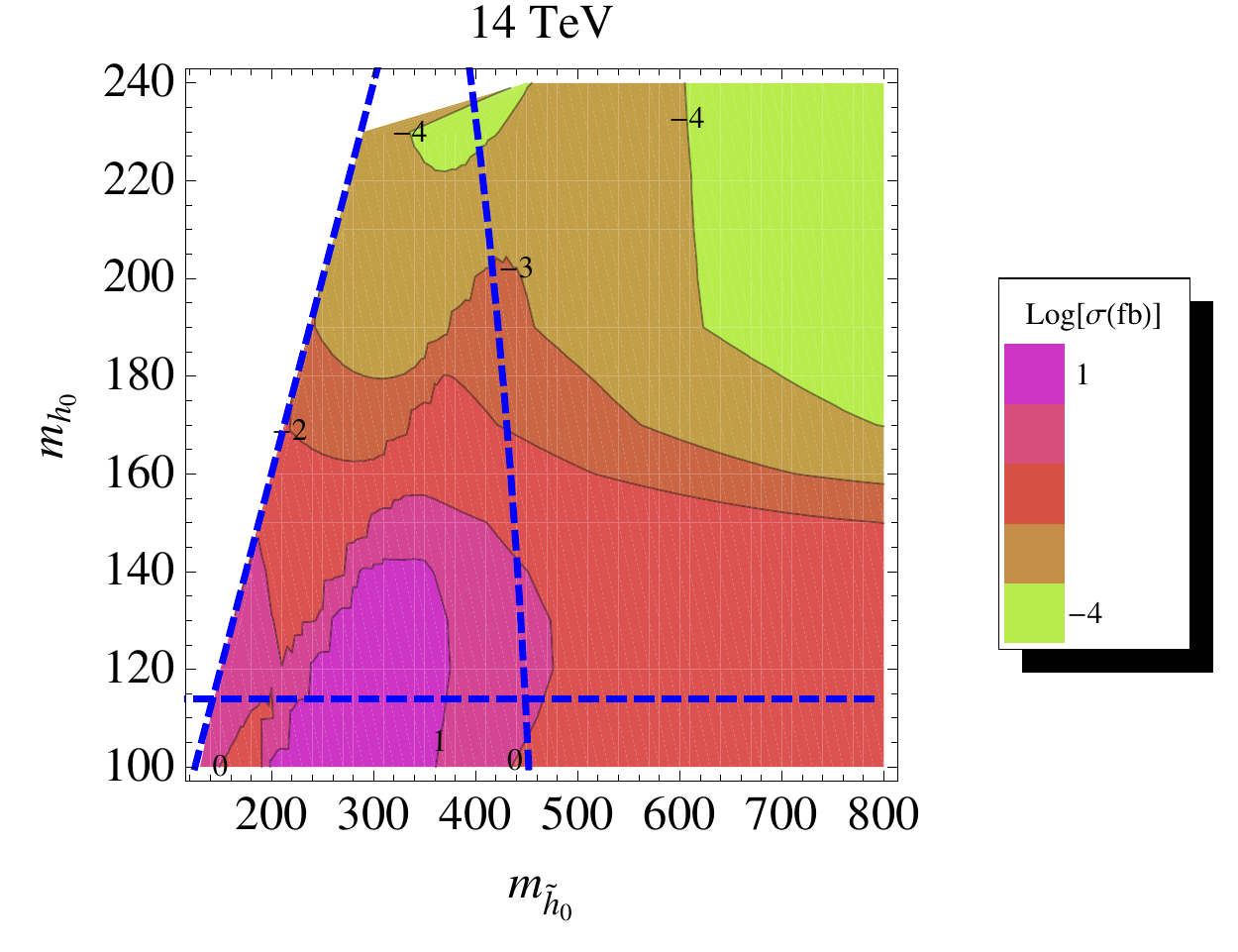}
}
\caption{Contour plot of the total cross section (in fb) for $gg \to h_{0} h_{0} \to b \bar{b} \gamma \gamma$ ($ \sqrt{S} = 7/14$ TeV respectively) versus the light Higgs boson mass, $m_{h}$ and 
heavy Higgs boson mass, $m_{\tilde{h}}$ for $M_{Q}= 0.5$ TeV and $M_{1} = M_{2} =1$ TeV. 
}
\label{fig:contour2}
\end{figure} 

Shown in figure~\ref{fig:contour2} are scans at both $\sqrt{S}=7/14~{\rm TeV}$ respectively
of the cross section $\sigma(h_0 h_0 \to \gamma \gamma b \bar{b})$ over the plane of  $(m_{h_0},m_{\tilde h_0})$.
 At $\sqrt{S}=14$ TeV we choose  three benchmark points  listed in Table~\ref{tbl:delm}. At $7$ TeV,  $\sigma(h_0 h_0 \to \gamma \gamma b \bar{b})$ is less than or close to $1~{\rm fb}$ throughout the 
plane of $(m_{h_0},m_{\tilde h_0})$. This is before any sort of event selection which would 
reduce this by a factor of $10$. 
Bearing in mind that the $7$ TeV LHC is expected to accumulate about $10~{\rm fb^{-1}}$ of integrated luminosity 
before its upgrade to $14\,{\rm TeV}$ we do not follow $7\,{\rm TeV}$ any further. 

\begin{figure}
\centerline{ \includegraphics[scale=0.7]{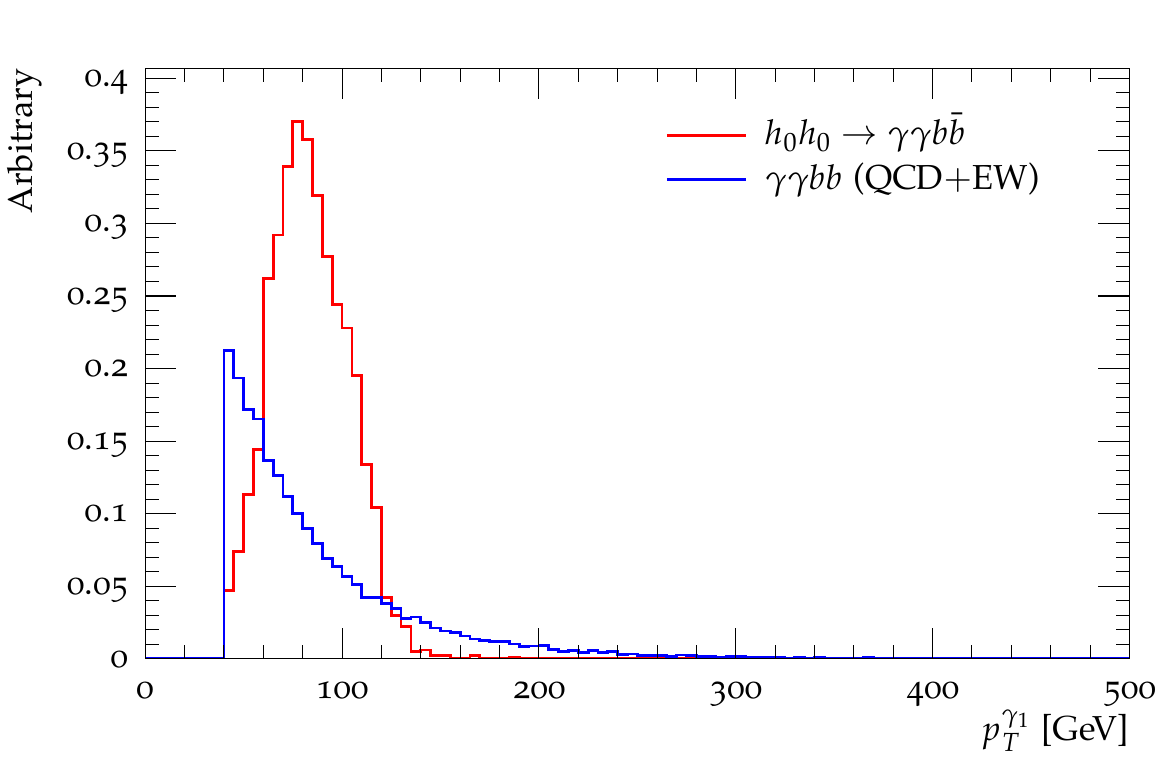} 
                   \includegraphics[scale=0.7]{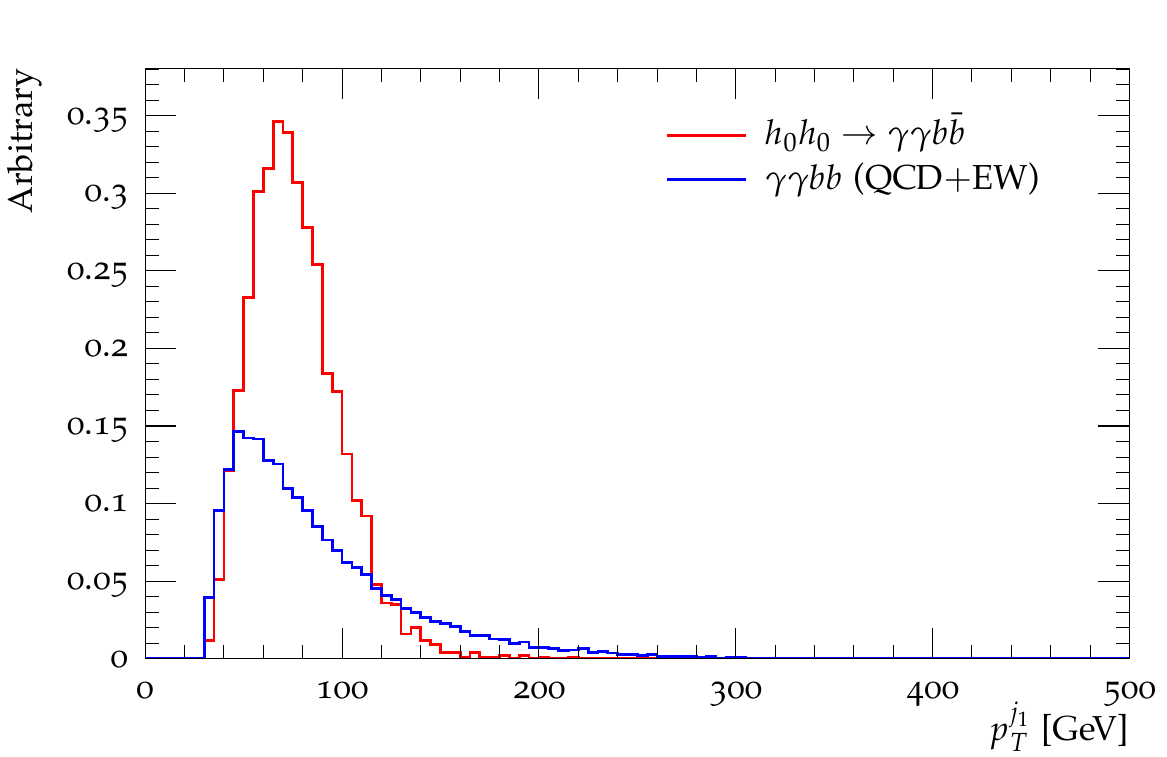}
  }
 \caption{Shown (in arbitrary units)  are the distributions for the signal process $h_{0} h_{0} \to \gamma \gamma b \bar{b}$ (red) and 
 one of the backgrounds, $\gamma \gamma bb$ (blue),
 in transverse momentum of the hardest jet $p_{T}^{j_{1}}$ (left) and hardest photon $p_{T}^{\gamma_{1}}$ (right).}
\label{fig:pt}
\end{figure}

At the LHC the signal process $pp \to \tilde{h}_{0} \to h_{0} h_{0} \to \gamma \gamma b \bar{b}$ will give rise to photons and jets of relatively high transverse momentum $p_{T} \sim 90~{\rm GeV}$.  In figure~\ref{fig:pt} we show the transverse momentum of the hardest photon and hardest jet to illustrate our point.
Backgrounds consist of (i) di-photon plus multi-jets, (ii) single photon plus multi-jets, and (iii) multi-jet production.  Our choice 
of photon isolation completely eliminates (iii) multi-jet production and (ii) single photon production from contention.  Out of di-photon plus multi-jets, the dominant 
contributions are from the associated production of two photons and two heavy flavours, i.e., bottom and charm quarks. These 
are denoted as $\gamma \gamma QQ$ where $Q=c,b,\bar{b},\bar{c}$.  In addition, there are backgrounds from $\gamma \gamma Qj$ and 
$\gamma \gamma jj$ where $j=u,d,s,g$. Photons and jets from these backgrounds tend to be softer than those from the our signal process (see figure~\ref{fig:pt}).

In our simulations we model $b$--tagging utilizing information in the event history of the Monte Carlo we are using.
We label a jet a $b$--tag if a partonic $b$-quark of at least $5$ GeV of transverse momentum is found in a cone of $R=0.3$ around the 
axis of the jet. If no $b$-quark is found, then we check in this order for a $c$-quark and $\tau$-lepton.  If no heavy quark or lepton is found, we 
label the jet a light jet.  Depending on which label the jet receives we apply the following weights: 
$\epsilon_{b}(E_{T},\eta)$, $\epsilon_{{\rm mistag}, c}=10 \%$, $\epsilon_{{\rm mistag},\tau}= 5 \%$, and 
$\epsilon_{{\rm mistag},j}= 0.5 \%$ \cite{Aad:2009wy,Altunkaynak:2010we}\footnote{The expression for 
$\epsilon_{b}(E_{T},\eta)$ is equal to the product of functions
$b_{E_{T}}$ and $b_{\eta}$. These functions are explicitly shown in ref.~\cite{Altunkaynak:2010we}.}, which is reflected in the results in table~\ref{tbl:cutflow}. 

For the computation of the backgrounds we have applied several parton--level cuts to regulate any soft or collinear divergences.
We require two $k_T$-jets with $D=0.7$ and 
\begin{eqnarray}
p_{T}^{\gamma} > 20~{\rm GeV}  \;,  \quad  p_{T}^{j} > 20~{\rm GeV} \; ,\\ \nonumber 
|\eta^{\gamma}| < 2.5 \;, \quad |\eta^{j}| < 2.5 \;, \quad R_{\gamma j} > 0.3 \;,  \quad R_{\gamma \gamma} > 0.3 \; .
\end{eqnarray}
For the signal process, $pp \to \tilde{h}_0 \to h_0 h_0 \to \gamma \gamma b \bar{b}$, we have not applied any parton--level cuts as 
there are no soft or collinear divergences.

We simulate events at the LHC using the Monte Carlo program {\tt Sherpa 1.3.0}
~\cite{Gleisberg:2008ta,Schumann:2007mg,Schonherr:2008av,Hoeche:2009rj}. 
We have  implemented the LWSM into {\tt Sherpa} and have subsequently generated matrix elements for 
$pp \to \tilde{h}_0 \to h_{0} h_{0} \to \gamma \gamma b \bar{b}$ using {\tt Amegic++}~\cite{Krauss:2001iv}.
The matrix elements for the background processes have been generated using {\tt Comix}~\cite{Gleisberg:2008fv}. 
All events generated include hadronization and shower effects. The parton shower is a Catani-Seymour subtraction based shower which is performed by
module {\tt CSSHOWER++}. Hadronization is performed by the module {\tt AHADIC++}. Additionally, the effects of soft QED radiation off hadron and tau decays has been simulated using the module {\tt PHOTONS++}.  

In order to analyze events we have written an analysis plugin for {\tt Rivet 1.3.0}~\cite{Buckley:2010ar}. {\tt Fastjet 2.4.2} has been used to perform the clustering of final state particles into jets~\cite{Cacciari:2005hq}. We have implemented the following selection criteria in our analysis:
\begin{itemize}
\item[Cut 1:]  
\begin{itemize}
\item
\emph{Photon isolation:}
i) $p_{T} > 20~{\rm GeV}$ ii) pseudo-rapidity range of $-2.5 < \eta_{\gamma} < 2.5$
are isolated photons if  iii) $\sum_{R \geq R_{\gamma k} }   E_{T}({k})<  0.1 p_T^\gamma$ is satisfied 
where $R_{\gamma k} \equiv \sqrt{(\phi_\gamma - \phi_k)^2 + (\eta_\gamma -\eta_k)^2 }$ and $R=0.3$. Here $k$ can be at the particle-level either 
hadrons or photons with $|\eta_{\gamma}|>2.5$ or $p_{T}^{\gamma} < 20~{\rm GeV}$.
\item  Exactly two isolated photons are required. 
\item The hardest isolated photon is required to have a minimal transverse momentum of $40~{\rm GeV}$ and  $R_{\gamma \gamma} >0.3$.
\end{itemize}

\item[Cut 2:]  Exactly two $k_{T}$-jets with $D=0.7$ in the pseudo-rapidity range of 
$-2.5 < \eta_j<2.5$ with minimal transverse momentum, $30~{\rm GeV}$, are required. 

\item [Cut 3:]  At least one $b$--tagged jet.

\item[Cut 4:] The di--photon invariant mass $M_{\gamma \gamma}$ is required to be in the mass window, $|M_{\gamma \gamma} - m_{h_0}| \le 2~{\rm GeV}$.

\item[Cut 5:] The dijet invariant mass $M_{b j}$  is required to be in the mass window, $|M_{bj} - m_{h_0}| \le 20~{\rm GeV}$.

\item[Cut 6:] The invariant mass $M_{bj \gamma \gamma}$ is required to be in the mass window, $|M_{b j \gamma \gamma} - m_{\tilde{h}_0}| \le \delta m_{\tilde{h}_0}$.
	            Values of our choice of $\delta m_{\tilde{h_0}}$ for each benchmark point are shown in Table~\ref{tbl:delm}.                   
\end{itemize}

\begin{figure}
\centerline{ 
\includegraphics[scale=0.7]{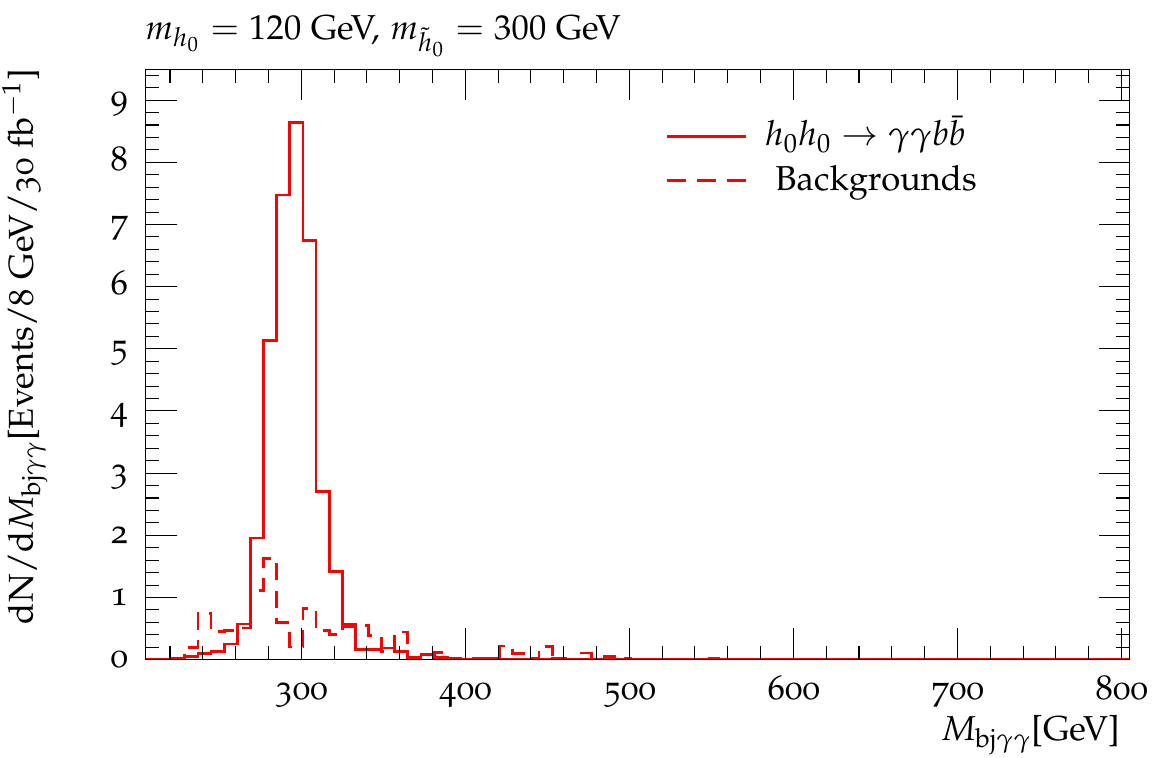}
  }
 \caption{Shown is the distribution in the invariant mass of two jets and two photons, $M_{bj \gamma \gamma}$, in $8$ GeV bins 
 for $30~{\rm fb}^{-1}$ of integrated luminosity at the $\sqrt{S}=14$ TeV LHC.}
\label{fig:mbjPP}
\end{figure}

\begin{table}
\begin{center}
\begin{tabular}{|cccc|}
\hline
Benchmark & $m_{h_0}$(GeV) & $m_{\tilde {h}_0}$(GeV)  & $\delta m_{\tilde{h}_0}$(GeV)  \\
\hline
(a) & 120 & 300 & 40 \\
(b) & 130 & 445 & 45 \\
(c) & 130 & 550 & 50 \\
\hline
\end{tabular}
\end{center}
\caption{Shown in this table are the light Higgs boson mass parameters $m_{h}$, the LW Higgs boson mass parameters, $m_{\tilde h}$, and the 
mass window parameters $\delta m_{\tilde{h}_0}$ for benchmark points (a),(b), and (c). }
\label{tbl:delm}
\end{table}

\begin{table}
\begin{center}
\begin{tabular}{|c|c|cccccc|}
\hline 
& {\tt QCD$+$EW}:        & $\gamma \gamma  jj$ 
                          & $\gamma \gamma bb$ 
                          & $ \gamma \gamma cc$ 
                          & $\gamma \gamma bc$ 
                          & $\gamma \gamma bj$ 
                          & $\gamma \gamma cj$   \\
\hline
& $\sigma_{\rm gen}$(pb) & $23.2$ & $0.176$ & $1.56$ & $0.0840$ & $0.519$ & $6.26$   \\
\hline
& cut 1                               & $0.390$ & $0.370$ & $0.306$ & $0.295$ & $0.344$ & $0.354$  \\
& cut 2                               & $0.363$ & $0.358$ & $0.386$ & $0.435$ & $0.406$ & $0.366$ \\
& cut 3                               & $0.0526$ & $0.795$ & $0.116$ & $0.516$ & $0.460$  & $0.0920$ \\
\hline
& cut 4a                            & $0.0212$ & $0.0233$ & $0.0247$ & $0.0217$ & $0.0240$  & $0.0200$  \\
& cut 5a                            & $0.249$ & $0.229$ & $0.232$ & $ 0.242$ & $0.264$  & $0.203$   \\
& cut 6a                            & $0.604$ & $0.547$ & $0.713$ & $0.534$ & $0.471$ & $0.627$  \\
\hline
& $\epsilon_{\rm tot}$ & $2.37 \times 10^{-5}$
                                &  $3.07 \times 10^{-4}$ 
                                & $5.60 \times 10^{-5}$ 
                                &  $1.85 \times 10^{-4}$ 
                                & $1.93 \times 10^{-4}$ 
                                & $3.03 \times 10^{-5}$  \\
(a) & $\sigma_{\rm eff}$(fb) & $0.550$ & $ 0.0527$ & $0.0873$ & $0.0156$ & $0.100$ & $0.190$  \\
\hline
\hline
& cut 4b                               & $0.0150$ & $0.0202$ & $0.0139 $ & $0.0167$ & $0.0221$  & $0.0191$  \\
& cut 5b                               & $0.221$ & $0.213$ & $0.174$ & $ 0.242$ & $0.234$  & $0.276$   \\
& cut 6b                               & $0.136$ & $0.0567$ & $0.129$ & $0.138$ & $0.165$ & $0.130$ \\
\hline
& $\epsilon_{\rm tot}$ & $3.37 \times 10^{-6}$
                               &  $2.56 \times 10^{-5}$ 
                               & $6.14 \times 10^{-6}$ 
                               &  $3.67 \times 10^{-5}$ 
                               & $5.46 \times 10^{-5}$ 
                               & $8.06 \times 10^{-6}$  \\
(b) & $\sigma_{\rm eff}$(fb) & $0.0782$ & $0.00431$ & $0.00959$ & $ 0.00309$ & $0.0283$ & $0.0505$  \\
\hline
\hline
& cut 4c                               & $0.0150$   & $0.0213$  & $0.0199$     & $0.0167$     & $0.0221$   & $0.0191$  \\
& cut 5c                               & $0.221$     & $0.213$    & $0.174$       & $ 0.242$      & $0.234$     & $0.274$   \\
& cut 6c                              & $0.00723$ & $0.0337$   & $0.00289$   & $0.0164$     & $0.0303.$   & $0.0.0122$  \\
\hline
& $\epsilon_{\rm tot}$       & $1.79 \times 10^{-7}$&  
                                         $1.52 \times 10^{-5}$ & 
                                         $1.38 \times 10^{-8}$ &  
                                         $4.36 \times 10^{-6}$ & 
                                         $1.00 \times 10^{-5}$ & 
                                         $7.58 \times 10^{-7}$  \\
(c) & $\sigma_{\rm eff}$(fb) & $0.00414$ & 
                                       $ 0.00261$ & 
                                       $2.15 \times 10^{-5}$ & 
                                       $0.000366$ & 
                                        $0.00521$ & 
                                        $0.00475$   \\
\hline
\end{tabular}
\end{center}
\caption{
Table of cross sections (in pb) for benchmarks (a),(b), and (c) before selection cuts ($\sigma_{\rm gen}$) and with selection cuts ($\sigma_{\rm eff}$) for 
the backgrounds $QQ\gamma \gamma$, $Qj\gamma\gamma$, and $jj\gamma\gamma$ where $Q=c,b,\bar{c},\bar{b}$ and $j=u,\bar{u},d,\bar{d},s,\bar{s},g$
for $\sqrt{S}=14$ TeV. Efficiencies (cuts $1$--$6$) are relative where $\epsilon_{tot}$ is the cumulative efficiency. Cuts 1-3 are reproduced only once as they are the same for 
all three benchmarks.}
\label{tbl:cutflow}
\end{table}
%

\begin{table}
\begin{center}
\begin{tabular}{|c||c||c||c|}
\hline
$pp \to h_{0} h_{0} \to \gamma \gamma b \bar{b}$   & 
(a)  & (b) & (c) \\
\hline 
     $\sigma_{\rm gen}$(fb)  & $11.2$ &$0.964$ & $0.195$ \\
     \hline 
     cut 1 & $0.594$ & $0.675$ & $0.693$ \\
     cut 2 & $0.414$ & $0.405$ & $0.391$ \\
     cut 3 & $0.734$ & $0.760$ & $0.748$ \\
     cut 4 & $0.999$ & $0.999$ & $0.999$ \\
     cut 5 & $0.601$ & $0.567$ & $ 0.586$\\
     cut 6 & $0.966$ & $0.823$ & $0.725$ \\
     \hline
     $\epsilon_{\rm tot}$ & $0.105$ & $0.097$ & $0.0861$ \\
     $\sigma_{\rm eff}$(fb)  & $1.18$ &  $0.0935$ & $0.0168$ \\
\hline
\end{tabular}
\end{center}
\caption{Cross sections (in fb) before selection and after selection for benchmarks (a) $m_{h_0}=120~{\rm GeV}$, $m_{\tilde h_0}=300~{\rm GeV}$, (b) $m_{h_0}=130~{\rm GeV}$, $m_{\tilde h_0}=445~{\rm GeV}$, and (c) $m_{h_0}=130~{\rm GeV}$, $m_{\tilde h_0}=550~{\rm GeV}$. Efficiencies (cuts $1$-$6$) are relative where $\epsilon_{\rm tot}$ is the cumulative efficiency.
}
\label{tbl:sigcutflow}
\end{table}

\begin{table}
\begin{center}
\begin{tabular}{|c|c|}
\hline
$pp \to h_{0} Z \to \gamma \gamma b \bar{b}$   & 
(a) $m_{h_0}=120~{\rm GeV}$, $m_{\tilde h_0}=300~{\rm GeV}$  \\
\hline 
     $\sigma_{\rm gen}$(fb)  &  $32.3$ \\
     \hline 
     cut 1  & $0.745$\\
     cut 2  &  $0.489$ \\
     cut 3  &  $0.772$\\
     cut 4  &  $0.999$\\
     cut 5  &  $0.184$ \\
     cut 6  &  $0.422$\\
     \hline
     $\epsilon_{\rm tot}$ & $0.0218$ \\
     $\sigma_{\rm eff}$(fb)  & $0.703$\\
     \hline
\end{tabular}
\end{center}
\caption{Cross sections (in fb) for $h_{0} Z \to \gamma \gamma b \bar{b}$ before selection and after selection for benchmark (a) $m_{h_0}=120~{\rm GeV}$, $m_{\tilde h_0}=300~{\rm GeV}$. 
Efficiencies (cuts $1$-$6$) are relative where $\epsilon_{\rm tot}$ is the cumulative efficiency.
}
\label{tbl:Zhcut}
\end{table}

\begin{figure}
\centerline{
 \includegraphics[scale=0.7]{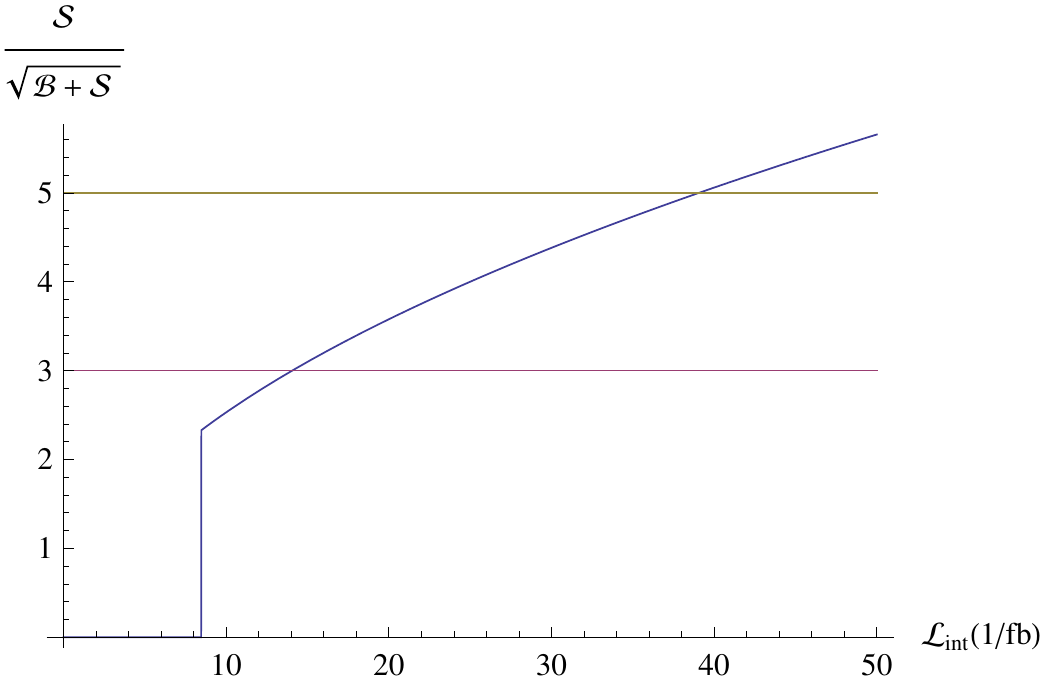} 
  \includegraphics[scale=0.7]{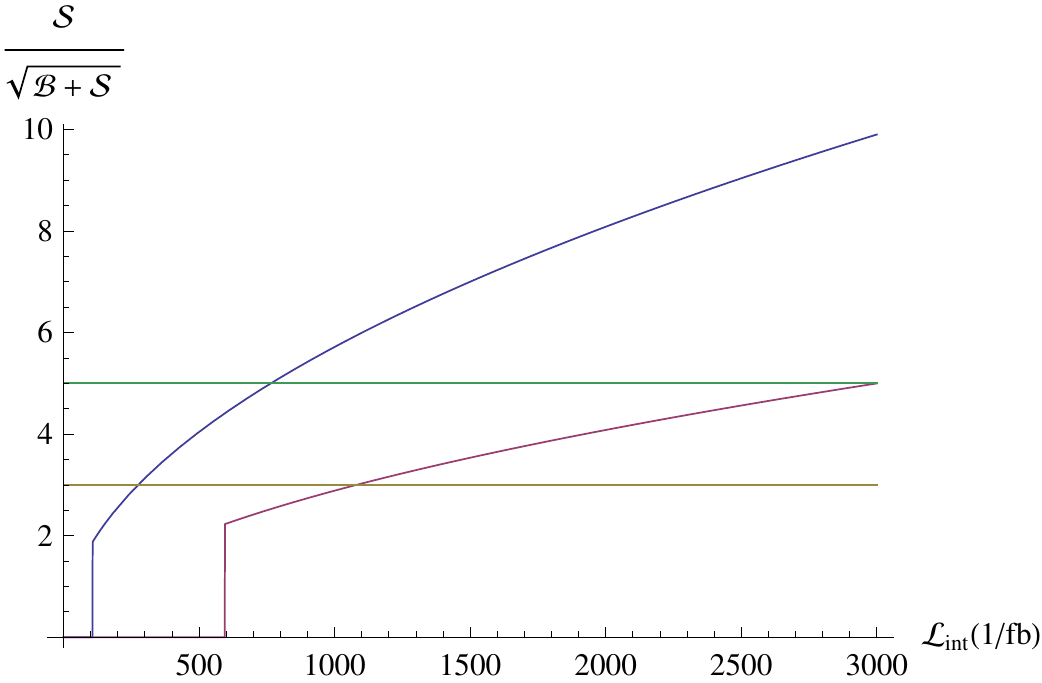}
}
\caption{Shown is the significance $\mathcal{S}/\sqrt{\mathcal{B}+\mathcal{S}}$ plotted against luminosity for benchmarks (a) (left) and benchmarks (b) in blue  and (c) in red (right). 
The upper and lower horizontal lines mark observation significances of $3\sigma$ and $5\sigma$. The vertical lines represent 10 events.}
\label{fig:sign}
\end{figure}

Table~\ref{tbl:cutflow} displays the efficiencies and cross sections for the backgrounds before and after selection cuts have been applied.  
Efficiencies and cross sections for the \emph{signal process} are shown in Table~\ref{tbl:sigcutflow}.  
In figure~\ref{fig:mbjPP} we show for $30~{\rm fb}^{-1}$ of integrated luminosity in $8$ GeV bins the invariant mass
of the $bj \gamma \gamma$ system for the signal scenario (a) and the sum of all backgrounds before cut $6$ has been applied.
For benchmark (a) we can expect to  establish a $5\sigma$-discovery with as little as $20~{\rm fb}^{-1}$.  
For benchmarks (b) and (c) outlook is not so optimistic. For scenario (b) we expect to reach $5\sigma$ at $700~{\rm fb}^{-1}$ and for
scenario (c) we would need $3000~{\rm fb}^{-1}$ of integrated luminosity. The primary reason for the reduced cross sections for scenarios (b) and (c) is that the dominant decay mode for the heavy LW Higgs $\tilde{h}_{0}$ is $\tilde{h}_0 \to \bar{t} t$ with
${\rm Br}_{\tilde{h}_{0}} \sim 95 \%$.

To this end we would like to mention that for benchmark (a)  there is a background from $Zh_{0}$ production\footnote{Note that 
benchmarks (b) and (c) this channel is dominated by top pairs.}. 
 Efficiencies and cross sections are shown in table~\ref{tbl:Zhcut}. It is worth mentioning that 
our analysis can be adapted for this case be changing our mass reconstruction hypothesis slightly. Instead of requiring the invariant mass
$M_{bj}$ to be in mass window around the $h_0$, we would instead, stipulate that in be in a mass window around the $Z$ boson. Additionally,
the invariant mass $M_{\gamma \gamma b j}$ should reconstruct the $\tilde{p}_{0}$.   

\section{Conclusions}
\label{sec:end}

In this paper we have investigated the possibility of a light LW Higgs sector.
As mentioned in the introduction SM-like Higgs sectors, such as the one of the LWSM, 
are not yet very well constrained as the the Higgs enters one-loop correction only logarithmically 
for larger masses and couples only very weakly to leptons
obscuring the clean di-lepton detection channel.
In practice this means that although the LW gauge bosons and the LW fermions
are constrained to lay in the few-TeV range the Higgs sector could be very low. 
In view of indirect (EWPO) and direct (collider) constraints we have assumed the 
SM-like Higgs boson to be below then $150 \,{\rm GeV}$-value. 

We have investigated such a possibility by looking at the cross sections
$gg \to h_0 h_0$ and $gg \to \tilde p_0 h_0$ c.f. figures~\ref{fig:contour},\ref{fig:contour1}  and the spectrum of $gg \to \bar tt$ figure~\ref{fig:mtt_histogram}.
Whereas the $gg \to h_0 h_0$  channel is outside reach at the LHC in the SM, it is enhanced   
in the LWSM in the case where the LW-like Higgs 
is twice as heavy as the SM-like Higgs ($m_{\tilde h_0} > 2 m_{h_0}$) and can decay at resonance through $gg \to \tilde h_0 \to h_0 h_0$ shown in figure~\ref{fig:gghh}(a).
The pseudoscalar $gg \to  \tilde p_0 \to \tilde p_0 h_0$ subprocess is 
close but not at resonance and turns out to be large as compared to SM Higgs channel but 
much smaller than the case discussed above as can be inferred from figure~\ref{fig:contour1} vs
\ref{fig:contour}. In our signal analysis we have therefore focused on the latter through 
$gg \to h_0 h_0 \to \bar b b \gamma \gamma$ and from table~\ref{tbl:sigcutflow} we see that 
the benchmark points (a) to (c) $(m_{h_0}, m_{\tilde h_0}) = \{(120,300),(130,445),(130,550)\}\,{\rm GeV}$ reach 10 events for integrated luminosities of 
$\{8.5,107 ,595\}\,{\rm fb}^{-1}$ and the $5\sigma$-discovery for $\{20,700 ,3000\}\,{\rm fb}^{-1}$ as can be seen from figure~\ref{fig:sign}.
In regard to these numbers we would like to add that the LHC is expected to collect $335 \,{\rm fb}^{-1}$ at $14\,{\rm TeV}$ from 2012 to 2020 before 
the upgrade to the Super LHC where
$1500\,{\rm fb}^{-1}$ is the reference number for 2025.

The Higgs pair production  cross section decreases rapidly for a $\tilde{h}_0$ with a mass above the top pair production threshold of $2m_{t}$.  In this region the intermediate states $\tilde h_0$ and $\tilde p_0$ decay mostly into top pairs as this is the dominant decay mode, c.f. figure~\ref{fig:branch}(right). In light of this it seems natural to investigate top pair production within the LWSM.  It is found though that the dip-peak or in general the visibility of the resonance is diluted when the width is large which happens when the intermediate states can decay into top pairs c.f. figure~\ref{fig:mtt_histogram}.
In the latter case the signal to background ratio can be significantly improved by 
applying $p_T$-cut of $250\,{\rm GeV}$ is applied to each top quark.  An example 
is given in figure~\ref{fig:mtt_histogram}(bottom-right) for 
$m_{h_0},m_{\tilde h_0} =  (125,800) \,{\rm GeV}$. Further suggestions on how to improve the 
signal are given in section \ref{sec:signal}.

\vspace{0.5cm}

Moreover, in this work we have also clarified a few things in the LWSM itself 
such as the tree-level sum rules in appendix \ref{app:masssumrules}, how to reduce 
hyperbolic diagonalizations to standard methods in appendix \ref{app:mass} and the issue
of spurious versus CP-violating phases in the LW generation Yukawa matrix in appendix~\ref{app:spurious}.
Moreover we have computed box diagrams with two vector (gluon) and pseudo/scalar (Higgs) 
flavour-changing vertex analytically, extending 
the results from the SM \cite{Glover:1987nx}  and MSSM  \cite{Plehn:1996wb}.\footnote{Flavour-changing vertices were computed in the MSSM 
in the squark sector \cite{Belyaev:1999mx} whereas here the top fermions are considered.} 
The results are  presented in appendix \ref{sec:box}.

%
%
\acknowledgments{We are grateful to Alexander Belyaev, Thomas Rizzo, Tilman Plehn,  Gustaaf Brooijmans, Rikkert Frederix, 
and Francesco Sannino for discussions.  RZ gratefully acknowledges 
the support of an advanced STFC fellowship. TF would like to thank the CERN Theory Division for their support. }
%
%
\appendix

\section{Results and definitions for $gg \to h_0h_0/ h_0 \tilde p_0$ process}
\label{app:gghh}

In this appendix all masses correspond to the physical masses and for 
the sake of notational simplicity  we shall use
the notation:
\begin{equation}
\mph{x} \to m_x
\end{equation}
for all the masses. We shall retain the subscript ${\rm phys}$ for the Yukawa matrices.
 
\subsection{Triangle graph}
\label{sec:triangle}
The triangle graph in the SM is given by\footnote{This notation agrees with 
\cite{Glover:1987nx} as follows: $a^\triangle_{0,2} = \text{gauge1(2)(triangle)}$.}:
\begin{equation}
{\cal A}^\triangle_0|_{\rm SM}(gg \to h_0 h_0)= \frac{-3m_H^2 s}{s-m_H^2+ i m_H \Gamma_H}  F_{1/2}(\beta_q)
\;, \quad \beta_{x} = 4 \mph{x}^2/s
\end{equation}
where
\begin{equation} 
F_{1/2}(x) = -2 x (1+(1-x)f(x))
\end{equation}
and 
\begin{equation}
\label{eq:f}
f(x) = \left\{ \begin{array}{ll}
{\rm Arcsin}^2(1/\sqrt{x}) & \quad x\geq 1  \\[0.2cm]
-\frac{1}{4}(\ln\big(\frac{1+\sqrt{1-x}}{1-\sqrt{1-x}}\big)-i \pi)^2 & \quad x < 1
\end{array} \right. \,.
\end{equation}
c.f.  \cite{Plehn:1996wb} for example\footnote{The function $f(x)$ relates to the Passarino-Veltman 
function as follows: $2 m_{x}^2/s \Big( 2 + (4 m_{x}^2 - s) C_{0}(0,s,0,m_{x}^2, m_{x}^2,m_{x}^2) \Big)=  \beta_{x} (1+(1-\beta_{x}) f(\beta_{x}))$.}

\subsubsection{$gg \to h_0/\tilde h_0 \to h_0 h_0$ triangles}
Since the Higgs sector \eqref{eq:hqq} does not contribute
to the loop, the LW-contribution can be obtained from the SM with modification 
of the vertices and taking into account mixing factors.
The coupling of the Higgs to the triangle itself is 
modified by mixing factors in eq.~\eqref{eq:s} $s_{H\!-\!\tilde H}$ and $\tilde s_{H\!-\!\tilde H} = - s_{H\!-\!\tilde H}$ for 
the standard and the LW Higgs boson respectively. The triple Higgs boson vertices
$h_0^3$ and $\tilde h_0 h_0^2$ are modified in the same way multiplying in addition 
a factor of $s_{H\!-\!\tilde H}^2$. Furthermore $\lambda v^2 =  2 \mph{ h_0}^2/(1+r_{h_0}^2)$,
exceptionally insisting on the subscript {\rm phys},  according to eq.~\eqref{eq:lambda} which leads to:
\begin{eqnarray}
\label{eq:th0h0}
 {\cal A}^\triangle_0(gg \to h_0 h_0)  =  
 \frac{-3 s_{H\!-\!\tilde H}^4 m_{h_0}^2}{1+r_{h_0}^2}  
 \Big( \frac{1}{s\! -\!m_{ h_0}^2 \!+\! i  m_{h_0}\Gamma_{h_0}}    
     -  \frac{1}{s\!-\! m_{\tilde h_0}^2\!-\! i  m_{\tilde h_0}\Gamma_{\tilde h_0}}  \Big) s~\tilde F_{1/2}
\end{eqnarray}
with
\begin{equation}
\label{eq:Ftilde}
\tilde  F_{1/2}   =
 \frac{( \physd{g}{t})_{11}}{m_{t}}     \,
F_{1/2}(\beta_{t}) -
  \frac{( \physd{g}{t})_{22}}{ {m_{\tilde t}}} \, F_{1/2}(\beta_{\tilde
t}) -
  \frac{( \physd{g}{t})_{33}}{ {m_{\tilde t'}}}   \,
F_{1/2}(\beta_{\tilde t'}) \,.
\end{equation}

%
%

\begin{figure}
 \centerline{
 \includegraphics{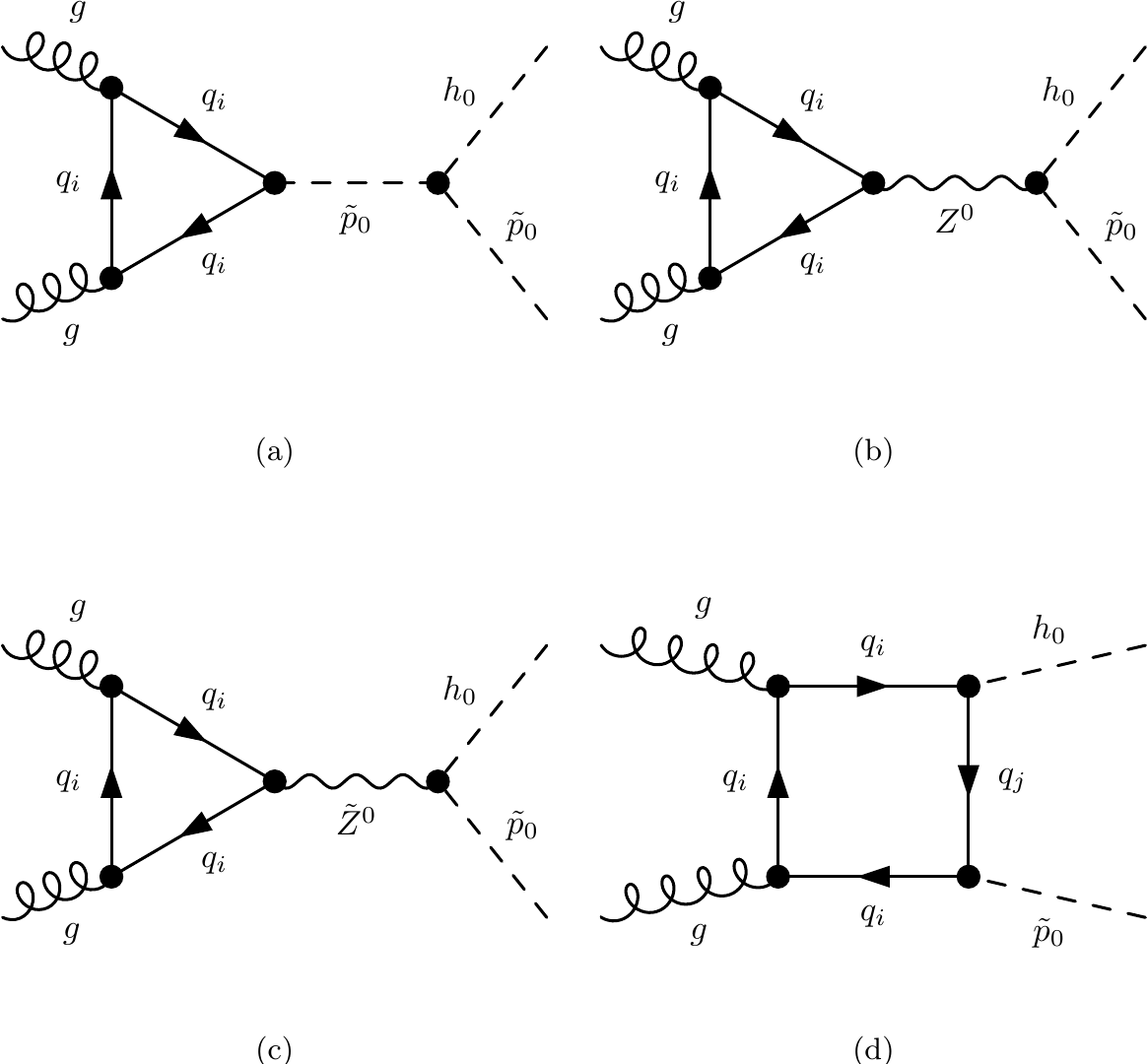}
}
\caption{(a--c) Triangle graphs for $q= (t, \tilde{t}, \tilde{T}, b,
  \tilde{b}, \tilde{B})$ and (d) one out of six box graphs for $q_{i}, q_{j}=(t,
  \tilde{t}, \tilde{T}, b, \tilde{b}, \tilde{B})$.}\label{fig:gghp}
\end{figure}

The process $gg \to h_0 \tilde p_0$ consist of triangles and boxes shown figure~\ref{fig:gghp}.  The triangle contributions can be broken down into contributions originating from:
\begin{enumerate}
\item $s$--channel $\tilde p_0$ exchange shown in figure~\ref{fig:gghp}(a),
\item $s$--channel $Z_0$ exchange shown in figure~\ref{fig:gghp}(b), and
\item $s$--channel $\tilde Z_0$ exchange shown in figure~\ref{fig:gghp}(c).
\end{enumerate}

We denote the contribution of all triangle diagrams by 
\begin{eqnarray}
{\cal A}^{\triangle}_{0}(gg \to h_0 \tilde p_0) &=& {\cal A}^{\triangle,\tilde p_0}_{0}  
+ {\cal A}^{\triangle, Z t }_{0}  + {\cal A}^{\triangle, Z b }_{0} \;,
\end{eqnarray}
where the amplitudes are further defined in the next subsection.

\subsubsection{$gg \to \tilde p_0 \to h_0 \tilde p_0$ triangles}

\begin{equation}
\label{eq:tp0h0} 
 {\cal A}^{\triangle,\tilde p_0}_{0}(gg \to \tilde p_0 h_0)   =  
  i \frac{s_{H\!-\!\tilde H}}{1+r_{h_0}^2}  \Big(  \frac{2  m_{h_0}^2 }{s-m_{ \tilde p_0}^2}  
  s~ \tilde P_{1/2} \Big)
\end{equation}
where for $\tilde P_{1/2} = \tilde F_{1/2}( F_{1/2}(\beta_x)  \to P_{1/2}(\beta_x))$ with 
$P_{1/2}(\beta_{x}) =  \beta_{x} f(\beta_{x})$ in accordance with \cite{Plehn:1996wb}. 

\subsubsection{$gg \to  Z_0/\tilde Z_0 \to h_0 \tilde p_0$ triangles}
\begin{eqnarray}
{\cal A}^{\triangle,Z q}_{0} &=& i \frac{e v^2 s_{\tilde H} (\cosh \theta_{Z} + \sinh \theta_{Z})}{\cos \theta_{W} \sin \theta_{W}}
(m_{\tilde p_0}^2 - m_{h_{0}}^2 ) \times \\[0.1cm]
& & \!\!\!\!\!\!
\sum_{j=1}^3 \eta_{jj} \left( 
 \frac{ (g_{R, {\rm phys}}^{Z q_j\bar{q_j}} - g_{L, {\rm phys}}^{Z q_j \bar{q_j}}) (1-\frac{s}{m_{Z}^2})}{s-m_{Z}^2 + i m_{Z}  \Gamma_{ Z}   }  -  \frac{ (g_{R, {\rm phys}}^{\tilde Z q_j\bar{q_j}} - g_{L, {\rm phys}}^{\tilde Z q_j \bar{q_j}}) (1-\frac{s}{m_{\tilde Z}^2})}{s-m_{\tilde Z}^2 - i m_{\tilde Z}  \Gamma_{ \tilde Z}   }  \right)
(1- \beta_{q_j} f(\beta_{q_j})) \nonumber
\end{eqnarray}
where the function $f$ is defined in \eqref{eq:f} and $s_{\tilde H} \equiv \sinh(\phi_h)$ 
in accord with our notation in eq.~\eqref{eq:s}. Note this is due to the fact that
prior to diagonalization only the $\tilde h_0 \tilde p_0 Z$-coupling but not 
the $ h_0 \tilde p_0 Z$-coupling is present.
The couplings of quarks to gauges bosons are parametrized as follows: 
\begin{eqnarray}
\mathcal{L} = \sum_{f=t,b}^{}  \Big( \bar{\Psi}_{L}^{f} g_{L}^{Zf \bar{f}} (\slashit{Z} + \slashit{\tilde Z}) \Psi_{L}^{f}  
       + \bar{\Psi}_{R}^{f} g_{R}^{Zf \bar{f}} (\slashit{Z} + \slashit{\tilde Z}) \Psi_{R}^{f}    \Big)_{\rm phys}         
\end{eqnarray}
The superscript ``phys'' indicates that all fields and couplings are understood to 
the physical ones. The physical couplings $g_{R,\rm phys}^{Zf \bar{f}}$ are obtained
from the expressions in Eqs. \eqref{eq:1} to \eqref{eq:4} as
$$g_{L, {\rm phys}}^X = S_{L}^{\dagger} g^X_{L} S_{L} \;, \qquad 
g^X_{R, {\rm phys}} = S_{R}^{\dagger} g^X_{R} S_{R} \;, ,$$
where $X$ stands for $Z t\bar t$ or $Z b \bar b$ respectively.
\begin{eqnarray}
\label{eq:1}
g_{R}^{Z t \bar{t}} = -\frac{e (\cosh \theta_{Z} + \sinh \theta_{Z})}{6 c_w s_w}
\begin{pmatrix}
-4(1 - c_w^2) & 0 & 0 \\
0 &4(1 - c_w^2) & 0 \\
 0 & 0 & -4c_w^2 + 1
\end{pmatrix}
\end{eqnarray}

\begin{eqnarray}
g_{L}^{Z t \bar{t}} = -\frac{e (\cosh \theta_{Z} + \sinh \theta_{Z})}{6 c_w s_w}
\begin{pmatrix}
4 c_w^2-1 & 0 & 0 \\
0 &4(1 - c_w^2) & 0 \\
 0 & 0 & -4c_w^2 + 1
\end{pmatrix}
\end{eqnarray}
\begin{eqnarray}
g_{R}^{Z b \bar{b}} = -\frac{e (\cosh \theta_{Z} + \sinh \theta_{Z})}{6 c_w s_w}
\begin{pmatrix}
2(1 - c_w^2) & 0 & 0 \\
0 & -2(1 - c_w^2) & 0 \\
0 & 0 & 2 c_w^2 + 1
\end{pmatrix}
\end{eqnarray}

\begin{eqnarray}
\label{eq:4}
g_{L}^{Z b \bar{b}} = -\frac{e (\cosh \theta_{Z} + \sinh \theta_{Z})}{6 c_w s_w}
\begin{pmatrix}
-2 c_w^2 - 1 & 0 & 0 \\
0 & -2(1 - c_w^2) & 0 \\
0 & 0 &  2 c_w^2 + 1
\end{pmatrix}
\end{eqnarray}

\subsection{Boxes for $gg \to h_0 h_0$  and $gg \to h_0 \tilde p_0$}
\label{sec:box}

For definiteness  we shall give one graph, the one indicated in figure~\ref{fig:gghh}(right):
\begin{eqnarray}
\label{eq:GF}
& &\Big[  (a_0)^\Box_{15})(m_i,m_j) (\tilde P_0)_{\mu\nu}  +  ( (a_2)^\Box_{15})(m_i,m_j)  (\tilde P_2)_{\mu\nu}  \Big]|_{\mbox{figure~\ref{fig:gghh}(right)}}  + X_{\mu\nu}  =    \nonumber \\[0.1cm]
& & (4 \pi^2 i) \int \frac{d^4 l}{(2\pi)^4} {\rm tr}[\gamma_\mu S_{m_i}(l+p_1) \gamma_\nu S_{m_i}(l+p_1 + p_2) {\mathbf 1} S_{m_j}(l+p_1+ p_2 + p_3) \gamma_5 S_{m_i}(l)]  \;, \nonumber 
\end{eqnarray}
for vertices $\mathbf{1}$ and $\gamma_5$. The term $X_{\mu\nu}$ stands for are structures vanishing 
when contracted with the according polarization vectors. 
As stated in the main text in this notation only $(a_{0,2})_{11}^{\Box}$ do contribute in the
SM, since there are no fundamental  pseudoscalars, 
and are related to the results in \cite{Glover:1987nx} as: 
$(a_{0,2})_{11}^{\Box} = \text{gauge}1(2)(box)$.

In the following we shall present our results for the box graphs. 
The analytic computations have been performed with the aid of 
FeynCalc \cite{Mertig:1990an}.
We are not aware of them being published elsewhere for the case where the flavour can change between the Higgs vertices.
The gluon momenta are  $p_1$ and $p_2$ whereas the  Higgs pair momenta are $p_3$ and are  $p_4$. We use the convention where all momenta are incoming, i.e.
$p_1  + p_2 =  - p_3 -  p_4$ . The result is given in terms of the Mandelstam variables 
\begin{equation}
s = (p_1+p_2)^2\;, \quad t = (p_1+p_3)^2 \;, \quad  u = (p_1+p_4)^2
\end{equation} 
and further shorthands
\begin{equation}
T_i = t - m_i^2  \;, \quad U_i = u - m_i^2  
\end{equation}
for $i = 3,4$.
\begin{eqnarray}
\label{eq:Aresults}
& &  \!\!\!\!\! \!\!   \!\!\!\!\! \!\! (a_0)^\Box_{11}(m,M)  \nonumber \\[0.1cm]  
&=&  \frac{1}{s} \Big\{ 4s   + 8 M^2 s C_{12} + 2 s ( (m+M)( 2 M^2(m+M)- M s)- M^2 (t+u)  )(D_{123}+D_{132}+D_{213}) \nonumber \\[0.1cm]  &+&
 (m_3^2+m_4^2 -2 (m+M)^2)\big[ T_3 C_{13} + T_4 C_{24} +
  U_3 C_{23} + U_4 C_{14} - 
 (t u - m_3^2 m_4^2 +s(m^2-M^2)) D_{132} \big]  \nonumber \\[0.1cm]  &+&  \{ m \leftrightarrow M \} 
\Big\}  \nonumber
 \end{eqnarray}
 \begin{eqnarray}   
 & & \!\!\!\!\! \!\!   \!\!\!\!\! \!\!    (a_0)^\Box_{51}(m,M) \nonumber \\[0.1cm]  
&=&  \frac{(-i)}{s} \Big\{  - 2 s(m M s + M^2(m_3^2 - m_4^2))(D_{123}+D_{132}+D_{213})
 \nonumber \\[0.1cm]  &+&   (m_3^2 - m_4^3)\big[ T_3 C_{13} + T_4 C_{24} +
  U_3 C_{23} + U_4 C_{14} - 
 (t u - m_3^2 m_4^2 +s(m^2-M^2)) D_{132} \big]  \nonumber \\[0.1cm]  &+&   \{ m \leftrightarrow M \} 
\Big\}  \nonumber \end{eqnarray}
 \begin{eqnarray}    
& & \!\!\!\!\! \!\!   \!\!\!\!\! \!\!    (a_0)^\Box_{55}(m,M)  =  - (a_0)^\Box_{11}(m,-M) = 
- (a_0)^\Box_{11}(-m,M)   \nonumber \\[0.3cm]  
& & \!\!\!\!\! \!\!   \!\!\!\!\! \!\!    (a_2)^\Box_{11}(m,M)  \nonumber  \\ 
&=&  \frac{1}{t u  - m_3^2 m_4^2 } \Big\{   (t^2 + u^2 - (4m^2+4 m M)(t+u) +4 (m-M)(m+M)^3+ 2 m_3^2m_4^2)   s C_{12} \nonumber \\[0.1cm]  &+&  
( m_3^2 m_4^2 + t^2 - 2t(m+M)^2)(T_3 C_{13} + T_4 C_{24} - st D_{213})\nonumber \\[0.1cm]  &+&  
( m_3^2 m_4^2 + u^2  - 2u(m+M)^2)(U_3 C_{23} + U_4 C_{14} - su D_{123})\nonumber \\[0.1cm]  &-&  (t^2 + u^2- 2 m_3^2 m_4^2)(t+u- 2(m+M)^2)C_{34} \nonumber \\[0.1cm]  &-& 
(t+u- 2(m+M)^2) ( (t u  - m_3^2 m_4^2)(m^2+M^2) + s (m^2-M^2)^2)(D_{123}+D_{132}+D_{213})
\Big\}    
\nonumber \\[0.1cm]  &+&  
(M^2-m^2)( 2 (m+M)^2 (u (2 s +t) - m_3^2 m_4^2)- m_3^2 m_4^2 (s-t-u)-t u (m_3^2+ m_4^2)- 2 s u^2) s D_{123} \nonumber \\[0.1cm]  &+& 
(M^2-m^2)( 2 (m+M)^2 (t (2 s +u) - m_3^2 m_4^2)- m_3^2 m_4^2 (s-t-u)-t u (m_3^2 +m_4^2)- 2 s t^2) s D_{213} \nonumber \\[0.1cm]  &+& 
  \{ m \leftrightarrow M \} 
 \nonumber  \end{eqnarray}
 \begin{eqnarray}  
& & \!\!\!\!\! \!\!   \!\!\!\!\! \!\!    (a_2)^\Box_{51}(m,M)  \nonumber  \\ 
&=&  \frac{-i}{t u  - m_3^2 m_4^2 } \Big\{  (2 (M^2-m^2)(u-t) -t^2 +u^2)  s C_{12} \nonumber \\[0.1cm]  &+&  
( m_3^2 m_4^2 - t^2 )(T_3 C_{13} + T_4 C_{24} - st D_{213})\nonumber \\[0.1cm]  &+&  
( m_3^2 m_4^2 - u^2 )(U_3 C_{23} + U_4 C_{14} - su D_{123})\nonumber \\[0.1cm]  &+&  (
(t + u)^2- 4 m_3^2 m_4^2)(t-u)C_{34} \nonumber \\[0.1cm]  &+& 
(t-u) ( (t u  - m_3^2 m_4^2)(m^2+M^2) + s (m^2-M^2)^2)(D_{123}+D_{132}+D_{213})
\Big\}    
\nonumber \\[0.1cm]  &+i&  
(M^2-m^2)((s-t+u)(t u - m_3^2 m_4^2) + 2 s u (u-t))  )  s D_{123} \nonumber \\[0.1cm]  &-i& 
(M^2-m^2)((s-u+t)(t u - m_3^2 m_4^2) + 2 s t (t-u))  ) s D_{213}  \nonumber \\[0.1cm]  &+& 
  \{ m \leftrightarrow M \} 
  \nonumber \\[0.3cm]  
& &  \!\!\!\!\! \!\!   \!\!\!\!\! \!\!   (a_2)^\Box_{55}(m,M) =  - (a_2)^\Box_{11}(m,-M) = -(a_2)^\Box_{11}(-m,M)
 \end{eqnarray}
 We would like to add three comment concerning symmetries in the amplitudes.
 First the relation, \begin{equation}
 \label{eq:symmi}
 (a_{0,2})^\Box_{55}(m,M) =  - (a_{0,2})^\Box_{11}(m,-M) = -(a_{0,2})^\Box_{11}(-m,M)
 \end{equation}
  follows from commuting the $\gamma_5$ from one
 pseudoscalar vertex to the other one. It is easy to see that doing  this is equivalent 
 to an overall factor of  $-1$ and changing all the masses in the nominators where the $\gamma_5$ passed from say $M \to -M$. This in turn is equivalent to eq.~\eqref{eq:symmi}.
Second, the amplitudes $(a_{0,2})^\Box_{15}(m,M)$ can be obtained from  $(a_{0,2})^\Box_{51}(m,M)$ 
by interchanging $p_3$ and $p_4$ which results in:
\begin{equation}
p_3 \leftrightarrow p_4 \quad  \Rightarrow  \quad 
 m_3 \leftrightarrow m_4 \;, u\leftrightarrow t\;,   C_{13}\leftrightarrow C_{14}\,,C_{23}\leftrightarrow C_{24}\,,D_{123}\leftrightarrow D_{213}
  \end{equation}
Thirdly the $a^{\Box}$ are manifestly symmetric under interchange of  
  $t$ and $u$. We note that  the matrix element without polzarization vectors contracted  
  is symmetric under interchange $(p_1,\mu) \leftrightarrow (p_2,\nu)$ which results in $t \leftrightarrow u$.
  Thus $(a)^{\Box} P_{\mu \nu}$ is symmetric and since $P_0$, $P_2$, $\tilde P_0$ ($\tilde P_2$) 
  are even (odd) respectively the same property holds for 
 $(a_0)^{\Box}_{(15/51)}$,  $(a_{0,2})^{\Box}_{(11/55)}$  ($(a_2)^{\Box}_{(15/51)}$)
 as can be seen from the formulae above.

\subsection{Tensor structures}
\label{sec:projectors}

The tensor structure for the parity-even case ${P_0,P_2}$ are given in  \cite{Glover:1987nx}:
\begin{eqnarray*}
S_z=0&:&\qquad P_0^{\mu\nu} =  g^{\mu\nu}-\frac{p_1^\nu p_2^\mu }{(p_1 p_2)} \\
S_z=2&:&\qquad P_2^{\mu \nu} = g^{\mu \nu}
               +\frac{p_3^2 p_1^\nu p_2^\mu}{p_T^2 (p_1 p_2)}
               -\frac{2 (p_2 p_3) p_1^\nu p_3^\mu}{p_T^2 (p_1 p_2)}
               -\frac{2 (p_1 p_3) p_2^\mu p_3^\nu}{p_T^2 (p_1 p_2)}
               +\frac{2 p_3^\mu p_3^\nu}{p_T^2} \;,
\end{eqnarray*}
whereas the one for the parity-odd case 
\cite{Plehn:1996wb} are:
\begin{eqnarray*}
S_z=0&:&\qquad \tilde P_0^{\mu\nu} = \frac{1}{(p_1 p_2)} \epsilon^{\mu \nu p_1 p_2} \\
S_z=2&:&\qquad \tilde P_2^{\mu \nu} = \frac{ p_3^\mu \epsilon^{\nu p_1 p_2 p_3}
               +p_3^\nu \epsilon^{\mu p_1 p_2 p_3}
               +(p_2 p_3) \epsilon^{\mu \nu p_1 p_3}
               +(p_1 p_3) \epsilon^{\mu \nu p_2 p_3} }
               {(p_1 p_2) p_T^2} \;,
\end{eqnarray*}
where  $ p_T^2 =2 (p_1p_3)(p_2p_3)/(p_1p_2)-p_3^2$
and the projectors $\{ P_0,\tilde P_0,P_2,\tilde P_2\}$ are  normalized as follows:
\begin{equation}
P_i \in \{ P_0,\tilde P_0,P_2,\tilde P_2\} \quad \text{s.t.} \quad P_i P_j = 2 \delta_{ij} \;.
\end{equation}
Note that there are two more structures with the properties of $\tilde P_0$ and on more
with the property of $\tilde P_2$. This is of no relevance as we have performed the
computation by contracting with helicity vectors.
The basis that we have chosen is
$p_1 = (p,0,0,p)$, $p_2 = (p,0,0,-p)$, $\epsilon(p_1,\pm) = \epsilon(p_2,\mp)= 1/\sqrt{2}(0,-1,\mp i,0)$ 
$p_3 = (\sqrt{m_3^2 + q^2},0,q \sin(\theta), q \cos(\theta))$ and 
$p_4 = (\sqrt{m_4^2 + q^2},0,-q \sin(\theta), -q \cos(\theta))$ where $q$ is determined through 
energy conservation $2p = \sqrt{m_3^2 + q^2} + \sqrt{m_4^2 + q^2}$.

\subsection{Passarino-Veltman functions}
\label{sec:int}
To present our results we use the standard Passarino-Veltman functions \cite{Passarino:1978jh}:
\begin{eqnarray}
& & C_{ij}(m_1,m_2,m_3)  =  \nonumber \\
& &  \qquad \int \frac{d^4k}{i \pi^2} \frac{1}{(k^2-m_1^2)((k+p_i)^2-m_2^2)((k+p_i+p_j)^2-m_3^2)} \\[0.1cm]
& & D_{ijk}(m_1,m_2,m_3,m_4) =   \nonumber \\
& & \qquad  \int \frac{d^4k}{i \pi^2} \frac{1}{(k^2-m_1^2)((k+p_i)^2-m_2^2)((k+p_i+p_j)^2-m_3^2)((k+p_i+p_j+p_k)^2-m_4^2)} \nonumber
\end{eqnarray}
and introduce the following abbreviations
\begin{alignat}{2}
& C_{12} \equiv C_{12}(M,M,M)    \qquad  & &    C_{13} \equiv C_{13}(M,M,m) \nonumber \\
& C_{14} \equiv C_{14}(M,M,m)    \qquad   & &    C_{23} \equiv C_{23}(M,M,m) \nonumber \\
& C_{24} \equiv C_{24}(M,M,m)    \qquad    & &    C_{34} \equiv C_{34}(M,M,m) \nonumber \\
& D_{123} \equiv D_{123}(M,M,M,m)  \qquad &   &    D_{132} \equiv D_{132}(M,M,m,m) \nonumber \\
& D_{213} \equiv D_{213}(M,M,M,m) \;.
\end{alignat}
The loss of information in the exact mass dependence of the $C$ and $D$ functions has to 
be taken into account when symmetrizing in $m$ and $M$ in formulae Eqs \eqref{eq:Aresults}.
\subsection{Additional plots}
\label{app:plots}
\begin{figure}[ht!]
 \centerline{
 \includegraphics[scale=0.7]{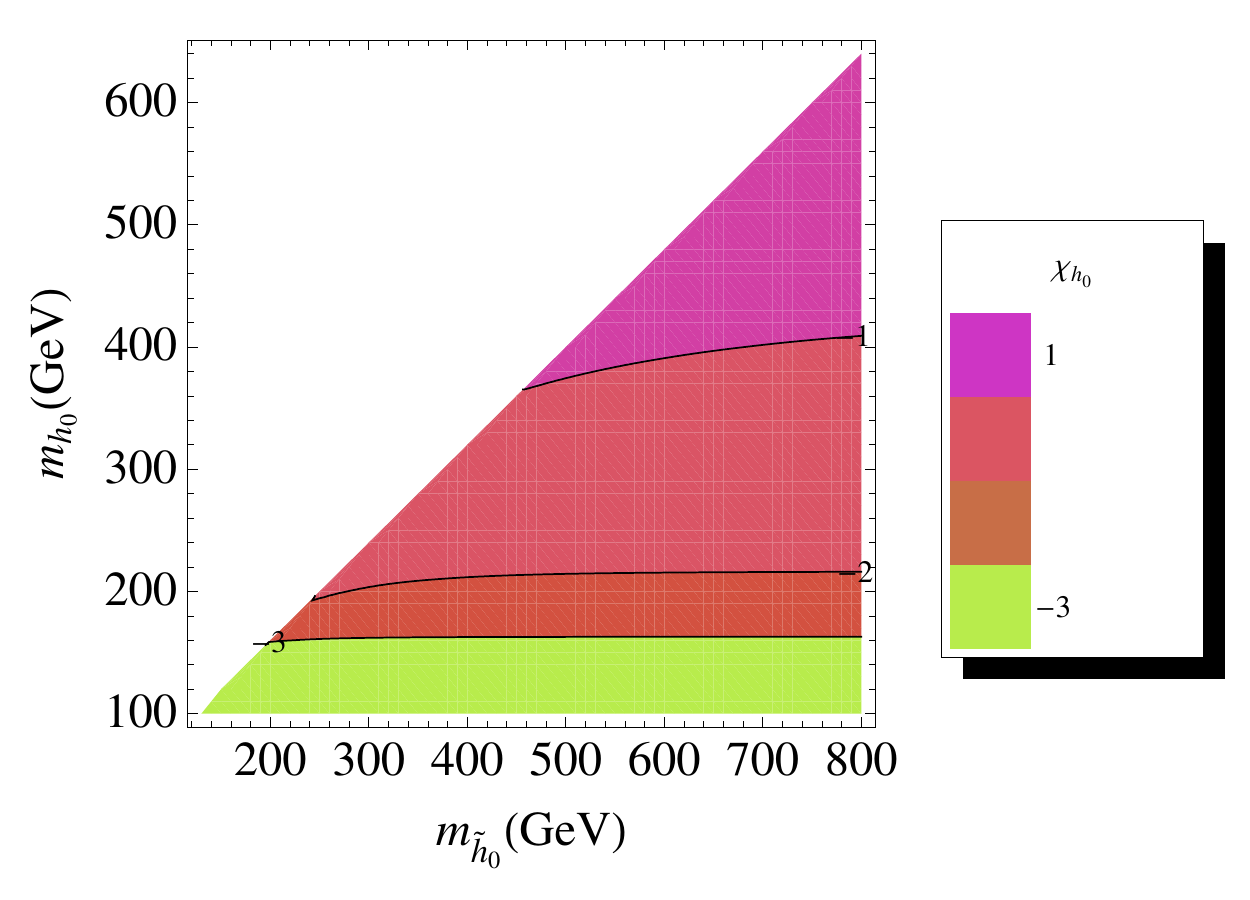}
 \includegraphics[scale=0.7]{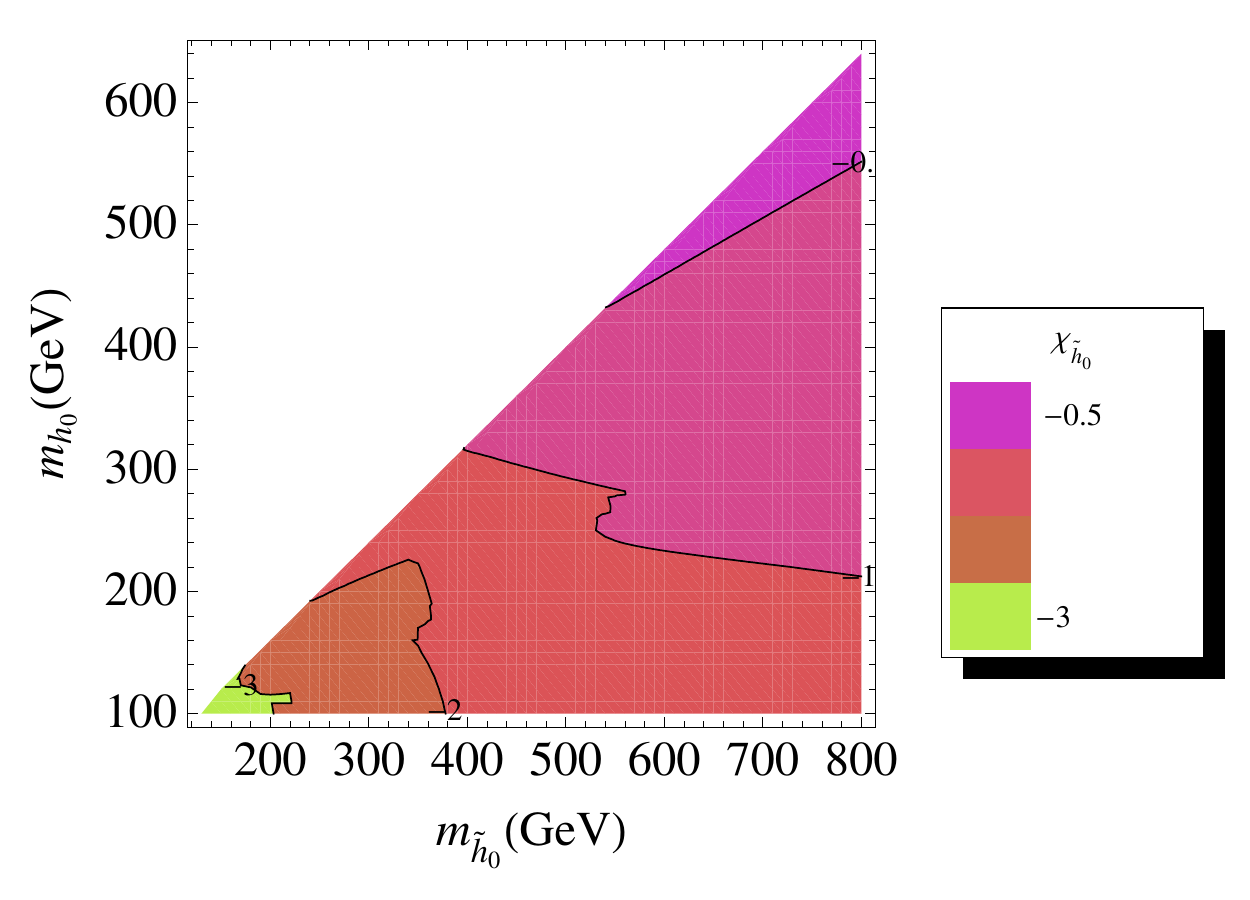}
}
\caption{Contours plots of the ratio $\chi_{h_0}=\log \Big( \frac{\Gamma_{h_0}}{m_{h_0}} \Big)$ (right) of the $h_0$ and 
the ratio $\chi_{\tilde h_0}=\log \Big( \frac{\Gamma_{\tilde h_0}}{m_{\tilde h_0}} \Big)$ of the $\tilde h_0$.
}
\label{fig:wcont}
\end{figure}

\begin{figure}[ht!]
 \centerline{
 \includegraphics[scale=0.7]{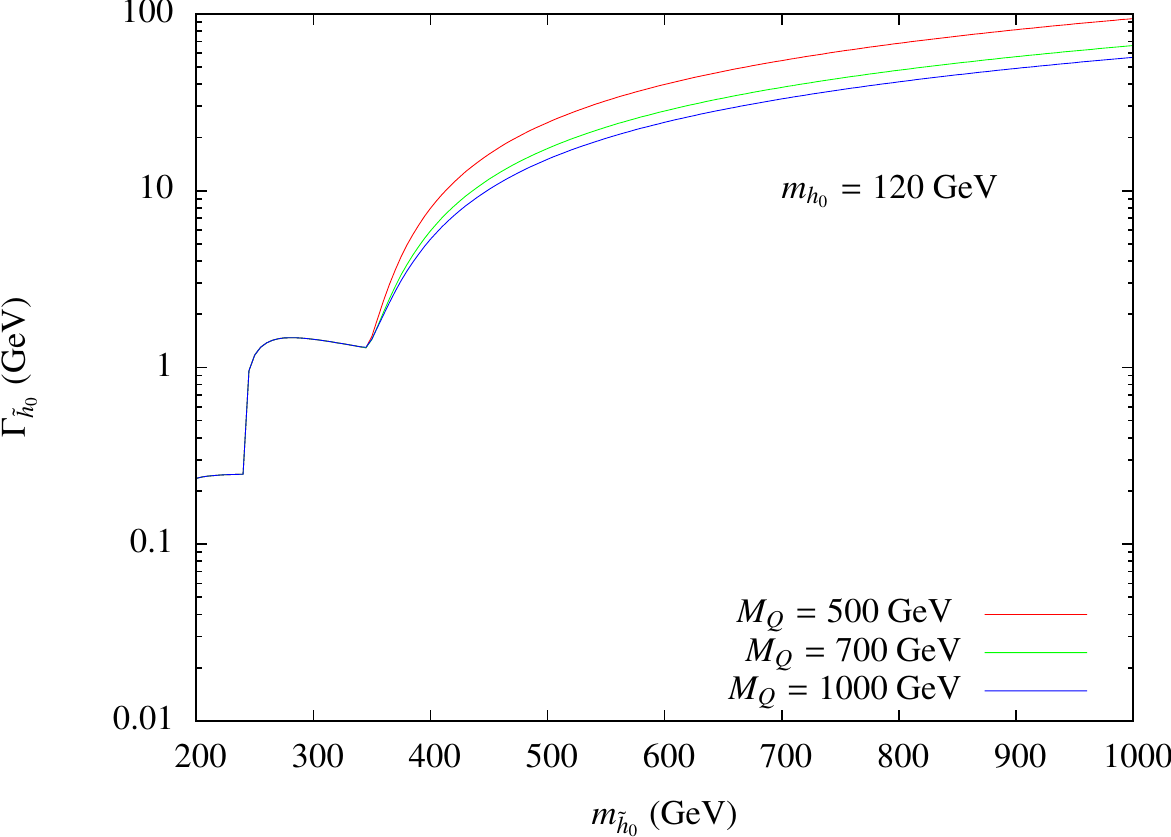}
 \includegraphics[scale=0.7]{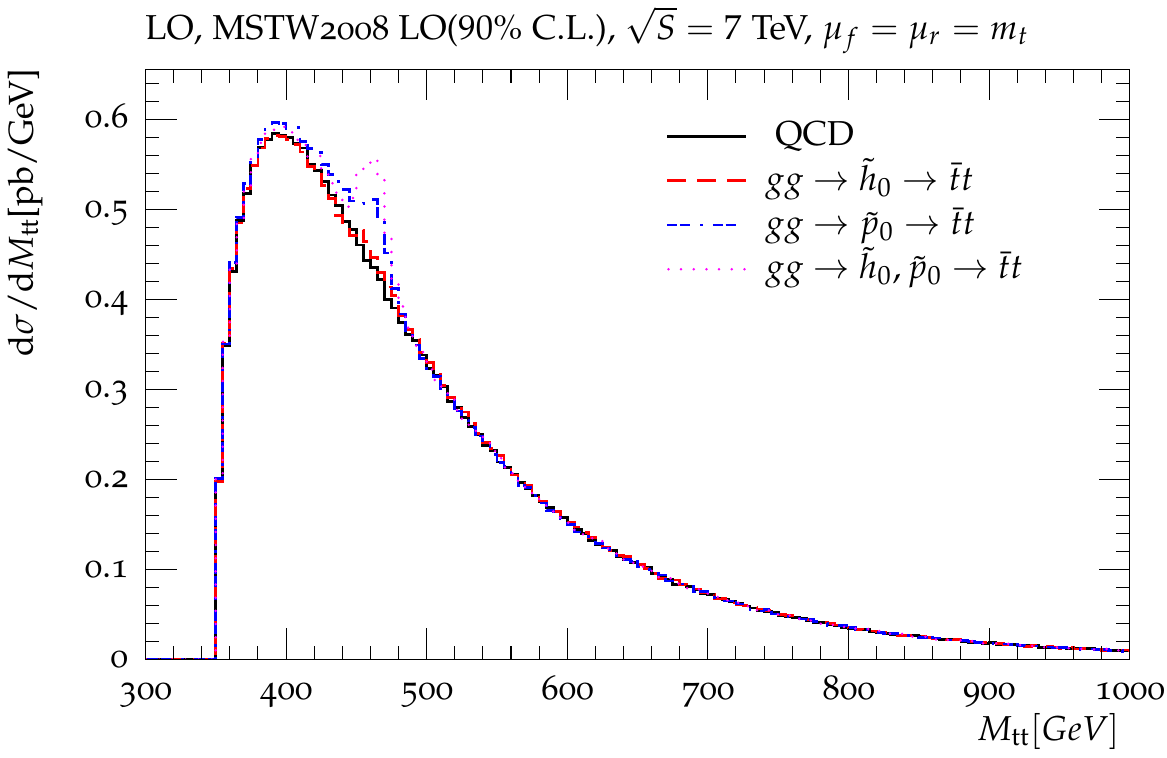}
}
\caption{(left) Width $\Gamma_{\tilde h_0}$  as a function of mass,
  $m_{\tilde h_0}$, for $m_{h_0} =120$ GeV, $M_{2}=M_{1}=1$ TeV 
  for different values of the fermion mass scale. (right) Histogram for $gg \to \bar tt$ for 
  $\sqrt{S} = 7\,{\rm TeV}$ with $5\,{\rm GeV}$-bins.}
\label{fig:width}
\end{figure}

\begin{figure}[ht!]
 \centerline{
 \includegraphics[scale=0.6]{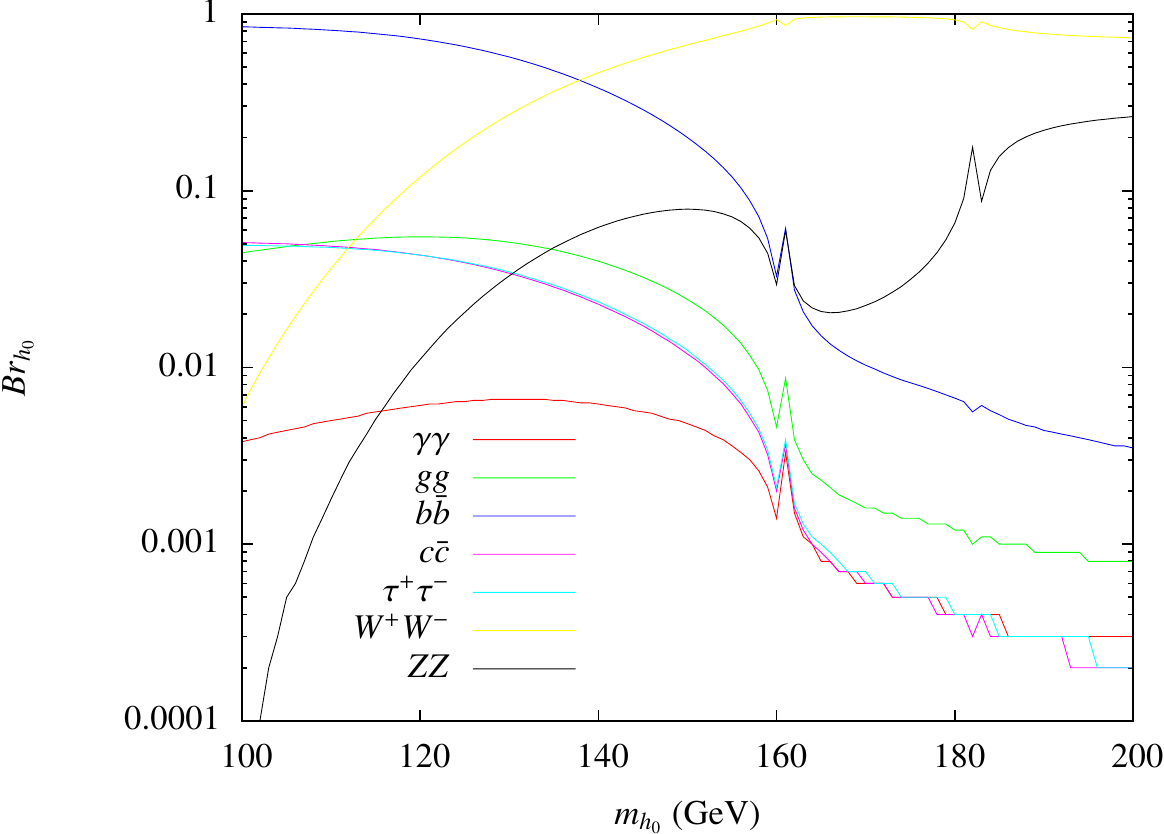}
  \includegraphics[scale=0.6]{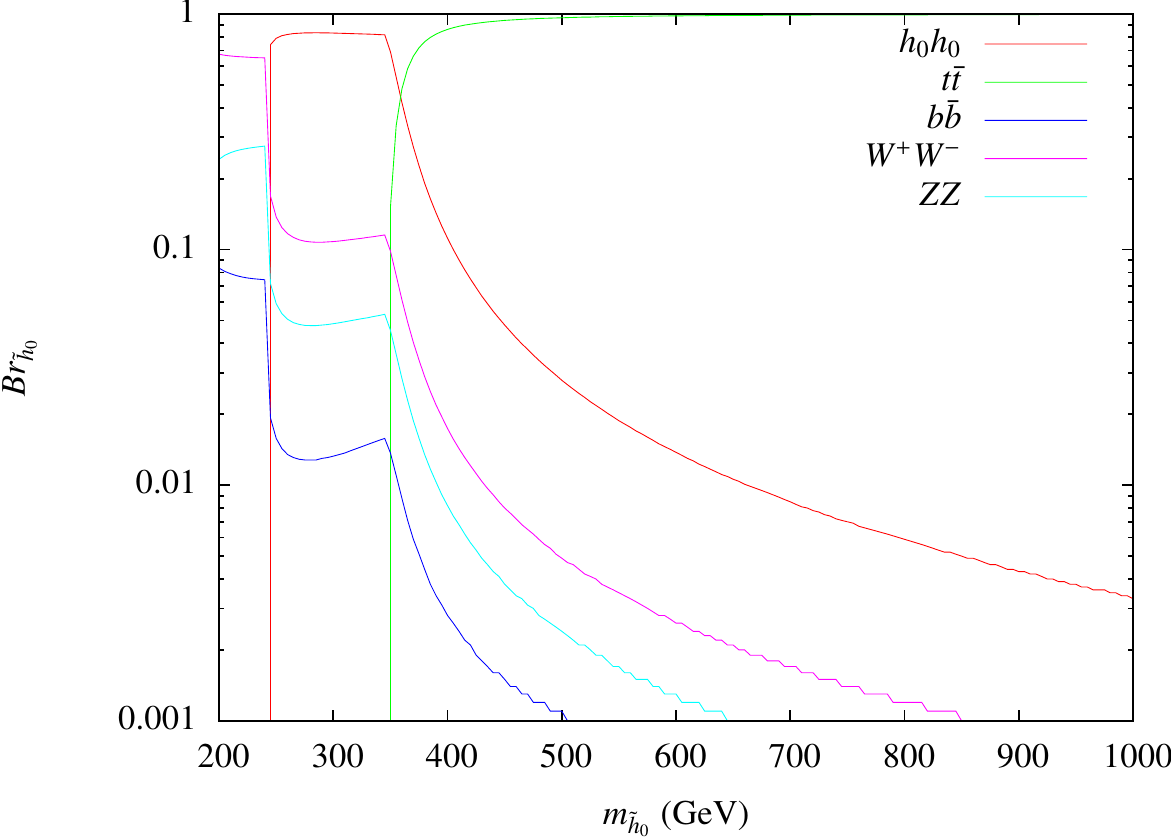} 
}
\caption{(left,right) Branching ratios ${\rm Br}_{h_0}$ and ${\rm Br}_{\tilde h_0}$  
as a function of 
  the masses $m_{h_0}$ and  $m_{\tilde h_0}$  and fixed $m_{\tilde h_0} =120$ GeV and 
   $m_{h_0} =450$ GeV respectively for $M_{2}=M_{1}=1$ TeV and $M_{Q}=500$ GeV.
  }
\label{fig:branch}
\end{figure}

\begin{figure}[ht!]
\centerline{
 \includegraphics[scale=0.6]{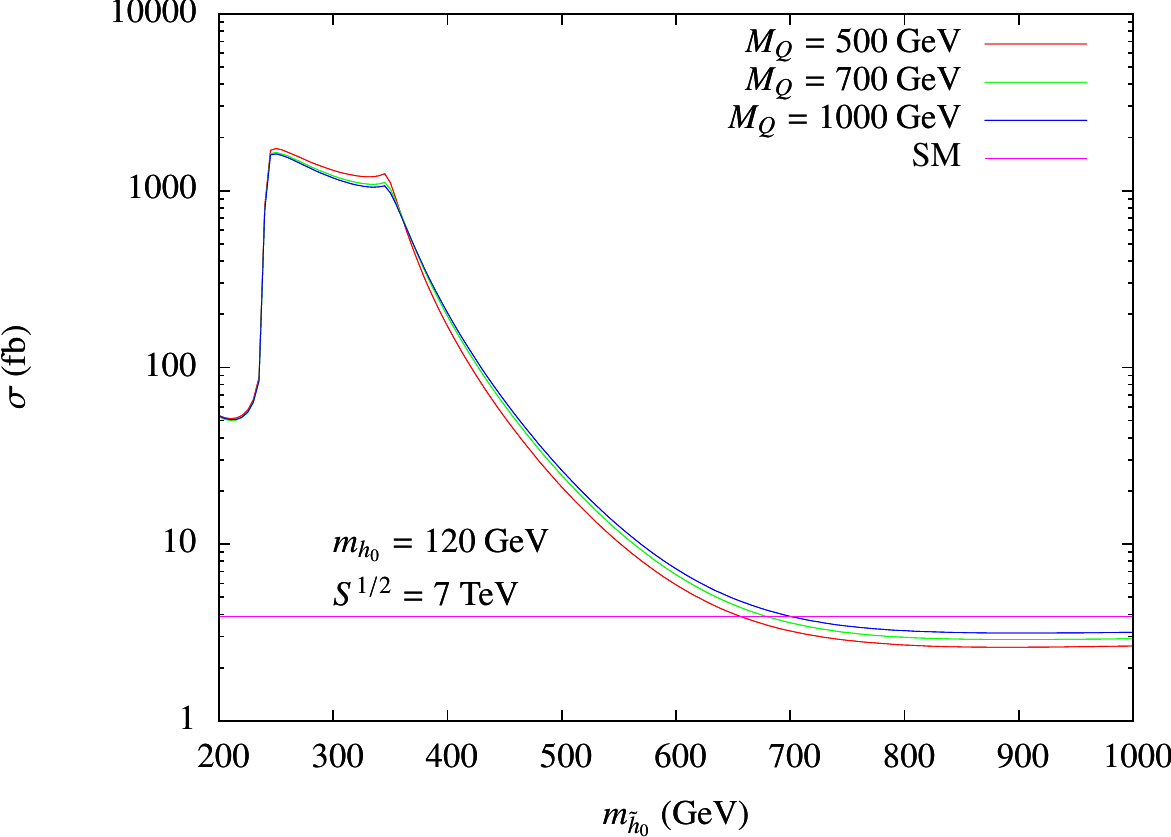} 
 \includegraphics[scale=0.6]{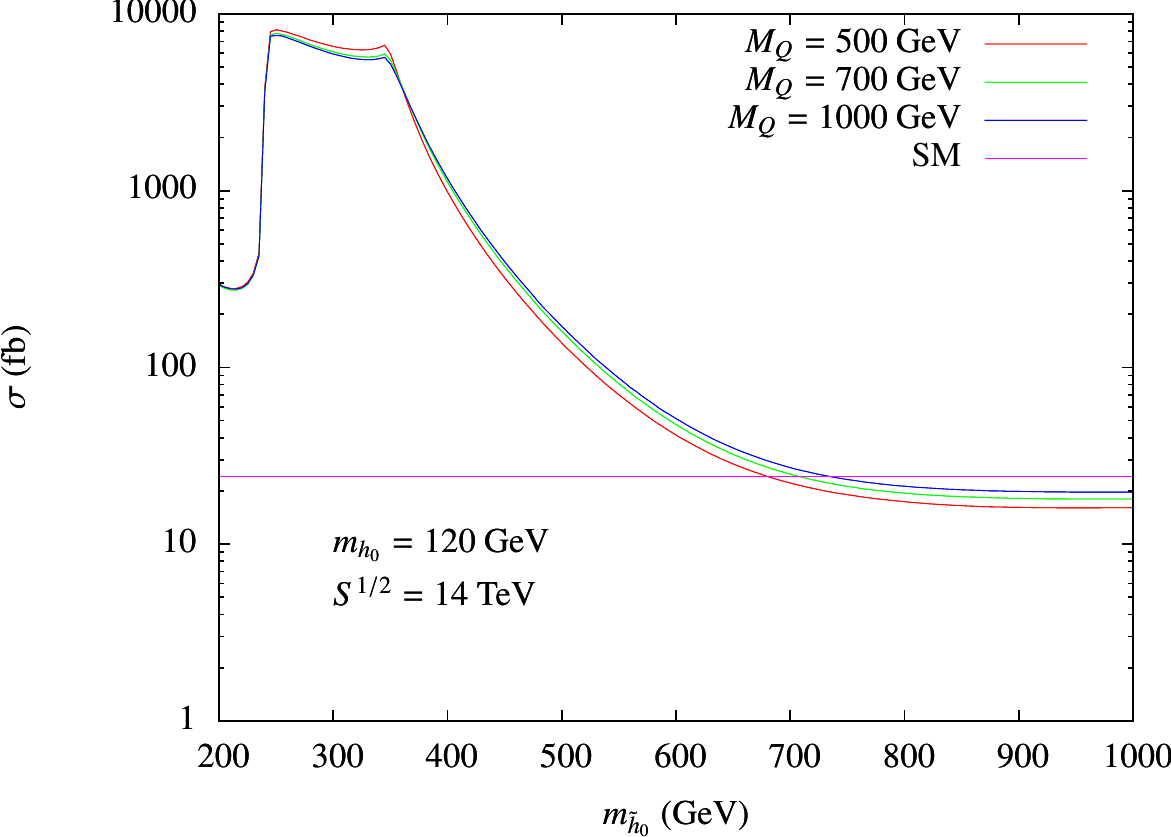}}
\caption{\small The cross section of $pp \to h_0 h_0$ via gluon fusion at
  the LHC for $\sqrt{s}=7/14$ TeV  respectively versus the mass of the $\tilde h_0$, $m_{\tilde h_0}$, for $m_{h_0} =120$ GeV.
  We note that the fermion mass scale $M_Q = M_u = M_d$ has very little influence 
  on the results as emphasized in section \ref{sec:numerical}. Note that for large $m_{\tilde h_0}$ the SM model value is approached by virtue of decoupling of the LW Higgs.}
\label{fig:mh2}
\end{figure}

\newpage
\section{Results for $gg \to h_0/\tilde h_0/ \tilde p_0 \to \bar tt$}
\label{app:ttresults}
The amplitudes for the processes can directly be obtained from the ones 
from the double Higgs pair production in the previous section by suitable replacements.
From the amplitude  $gg \to h_0 \to h_0 h_0$ in eq.~\eqref{eq:th0h0}, using 
eq.\eqref{eq:hqq} and the definition 
of $\lambda$ in the Higgs potential chosen in section \ref{sec:HiggsLW} one obtains:
\begin{eqnarray}
 {\cal A}^\triangle_0(gg \to h_0(\tilde h_0) \to  \bar t t )  = 
 {s_{H\!-\!\tilde H}^2  } (g^t_{\rm phys})_{11}
 \Big( \frac{1}{s\! -\!m_{ h_0}^2 \!+\! i  m_{h_0}\Gamma_{h_0}}    
     -  \frac{1}{s\!-\! m_{\tilde h_0}^2\!-\!i  m_{\tilde h_0}\Gamma_{\tilde h_0}}  \Big) s~\tilde F_{1/2} [\bar tt]
     \nonumber
\end{eqnarray}
Furthermore, from the  $gg \to \tilde p_0 \to \tilde p_0 h_0$ amplitude in eq.~\eqref{eq:tp0h0} 
one obtains:
\begin{equation}
 {\cal A}^{\triangle,\tilde p_0}_{0}(gg \to \tilde p_0  \to\bar t t )   =  
   \Big(  \frac{- 2 i  (g^t_{\rm phys})_{11} }{s-m_{ \tilde p_0}^2 - i m_{\tilde p_0} \Gamma_{\tilde p_0}}  
  s~ \tilde P_{1/2} \Big) [\bar t \gamma_5 t]
\end{equation}
Note, in both cases, we have not evaluated the spinors $t,\bar t$.

\section{Diagonalization of Mass Matrices}
\label{app:mass}

Here we shall describe a method for performing the hyperbolic diagonalization 
\begin{equation}
\label{eq:MSLSR}
\physd{\cal M}{t} \eta_3 \ =\ S_R^\dagger\,{\cal M}_t \eta_3 \,S_L 
\end{equation}
using similarity transformations for which standard tools, e.g.  {\tt Diag 1.3} \cite{Hahn:2006hr},
can be used, based on the observation that:
\begin{equation}
\label{eq:hey}
(S_{R/L} \eta_3)^{-1}  = S_{R/L}^\dagger \eta_3
\end{equation}
The latter relation is easily verified from eq.~\eqref{eq:SnSn}

Here we will describe a procedure of obtaining $S_{L}$ and $S_{R}$ numerically using routines provided. From there it is straightforward to verify that: 
First, we recognize that 
\begin{eqnarray}
\label{eq:M2diag}
{\rm diag}(  \mph{t}^2,   \mph{\tilde{t}}^{2} ,   \mph{\tilde{T}}^{2} ) &=& 
 \mathcal{M}_{t,{\rm ph}} \eta_{3} \mathcal{M}_{t,{\rm ph}}^{\dagger} \eta_{3}  \nonumber \\[0.1cm]  
  &=& A_R (  \eta_{3} \mathcal{M}_{t} \eta_{3} \mathcal{M}_{t}^{\dagger}) A_R ^{-1} 
  = A_L ( \eta_{3} \mathcal{M}_{t}^{\dagger} \eta_{3} \mathcal{M}_{t})  A_L ^{-1} \; .
\end{eqnarray} 
with 
$A_R \equiv S_{R}^{\dagger} \eta_{3}$ and $A_L \equiv \eta_{3} S_{L}^{\dagger}$.

\subsection{Mass sum rules}
\label{app:masssumrules}

In this section we would like to point out some tree-level sum rules for 
matrices. When the matrices are diagonalized by hyperbolic rotations the
trace remains an invariant. 
To be more precise suppose we had a matrix that is diagonalized as follows
\begin{equation}
\physs{{\cal M}}\,\eta = 
S ^\dagger\,{\cal M} \eta \,S \;, \qquad   
\end{equation}
with 
\begin{equation}
S^\dagger \eta S = \eta \;, \qquad  \physs{{\cal M}} = \text{diag}(\mph{a}^2, \mph{b}^2,...) \;,
\end{equation}
then 
\begin{equation}
\label{eq:trace}
\text{tr}[\physs{{\cal M}} ] = \text{tr}[{\cal M} ] \;.
\end{equation}
The correctness of \eqref{eq:trace} can be immediately verified using the properties above. 
The diagonalization can be interpreted as a symmetry transformation where 
$\eta$ plays the role of the metric. Thus the statement eq.~\eqref{eq:trace} is nothing
but the fact that the trace of the $(2,0)$-tensor $(M \eta)_{\alpha\beta}$ is an invariant;
 $ M_{\alpha}^{\;\;\alpha} =  \text{tr}[M]$. 
Thus one can deduce sum rules for the masses. Applied to 
the CP-even Higgs sector the  RHS follows from writing \eqref{eq:mh} in matrix form, c.f. 
 \cite{Krauss:2007bz} and the LHS is given by definition 
\begin{equation}
\label{eq:thiggs}
  \mph{h_0}^2 + m_{\tilde h_0}^2  =   M_H^2 = (\mph{\tilde p_0}^2 )  \;.
\end{equation}
The correctness is readily verified from eq.~\eqref{eq:mh}. Note, with the peculiar fact
that at tree-level $\mph{\tilde p_0}^2 = M_H^2$ equation \eqref{eq:tree} follows.
This technique applies to the entire bosonic sector.
For the neutral gauge bosons one gets 
\begin{equation}
\label{eq:tgbosons}
\mph{\tilde A}^2 + \mph{ Z}^2 + \mph{\tilde Z}^2  = M_1^2 + M_2^2
\end{equation}
with $M_{1,2}$ the mass scale of the $U(1)_Y$ and $SU(2)_L$  HD gauge terms respectively.
 The field $\tilde A$ is the LW-partner of the photon. Note the photon is not explicitly written down since it remains massless.
  eq.~\eqref{eq:tgbosons} is consistent with 
the result for $M_1 = M_2$ in appendix B of reference  \cite{Underwood:2008cr}.

The fermions are slightly more complicated as they proceed via a bi-unitary hyperbolic 
diagonalization. The statement is that:
\begin{equation}
\label{eq:tracef}
{\rm diag}(  \mph{t}^2, \mph{\tilde{t}}^{2} , \mph{\tilde{T}}^{2} ) \equiv \text{tr}[ \mathcal{M}_{t,{\rm ph}} \eta_{3} \mathcal{M}_{t,{\rm ph}}^{\dagger} \eta_{3}]  =
\text{tr}[ \mathcal{M}_t \eta_{3} \mathcal{M}_t^{\dagger} \eta_{3}] \;,
\end{equation}
which follows immediately from the eq.~\eqref{eq:M2diag}. Applied to 
the fermions we get:
\begin{equation}
\label{eq:tfermions}
 \mph{t}^2 + \mph{\tilde{t}}^{2} +  \mph{\tilde{T}}^{2} = M_u^2 + M_Q^2 \;,
\end{equation}
where eq.~\eqref{eq:Mmatrix} was invoked for $\mathcal{M}_t$. 
The correctness of this equation can be verified for the explicit result 
given in chapter 2.3.2. of reference  \cite{Krauss:2007bz} to each order in 
the expansion. In chapter 3 of reference \cite{Krauss:2007bz} similar consideration
were taken into account to show the absence of quadratic divergences in the 
top-loop in the AF formalism.

We would like to emphasize that the trace formula \eqref{eq:trace} and \eqref{eq:tracef} 
are general and in particular apply in each order of perturbation theory 
but the specific evaluation we have given in Eqs \eqref{eq:thiggs},\eqref{eq:tgbosons}
and \eqref{eq:tfermions} have made  use of the trace at tree-level 
and are thus subject to corrections.

\subsection{Spurious phases}
\label{app:spurious}
Furthermore we consider it worthwhile to discuss the freedom of reparametrizing
phases in the mass and Yukawa matrix of the LWSM.  Note that the Yukawa matrix
presented in ref.~\cite{Krauss:2007bz} contains imaginary entries and one might therefore
wonder whether they are associated with CP-violation or whether they are unphysical/spurious 
phases. 
For fixed flavour there are six fermion in each LW-generation counting left and right handed 
field separately. The freedom of choosing their spurious phases is reflected in 
the fact that the matrices $A_L$ and $A_R$ are determined by eq.\eqref{eq:M2diag} 
up to
\begin{equation}
A_R \to {\rm diag}(e^{i R_1},e^{i R_2},e^{i R_3}) A_R  \;, \quad 
A_L \to {\rm diag}(e^{i L_1},e^{i L_2},e^{i L_3}) A_L  
\end{equation}
a multiplicative diagonal unitary matrix.
Rewriting eq.\eqref{eq:MSLSR} as
\begin{equation}
\label{eq:MSLSR2}
\physd{\cal M}{t} \eta_3 \ = A_R (\eta_3{\cal M}_t \eta_3) A_L^{-1} 
\end{equation}
we see that choosing the fermion masses to be real and positive (or negative) fixes 
the differences $R_i - L_i$ for $i = 1,2,3$.  Writing 
$L_1 = L_1,  L_2 = L_1 + \Delta_2, L_3 = L_1 + \Delta_3$ it is noticed, as usual, that
only the two parameters   $\Delta_2$ and $\Delta_3$ lead to a change in the entries 
of $g_{t,\textrm{phys}}$; two arbitrary phases. This freedom can be used 
to reparametrize the third LW-generation by  $e^{i R_3} = e^{i L_3} = i$ the Yukawa matrix
 $g_{t,\textrm{phys}}$ in ref.~\cite{Krauss:2007bz} to render its entries completely real.

To this end we would like to note that we find that $g_{t,\textrm{phys}}$  
is smooth in the  limit $M_Q \to M_u$  
contrary to a remark made  in the appendix of ref.~\cite{EW1}.
Note in their explicit formula these authors present an expansion 
in $1/(M_u - M_Q)$ which cannot be compared with the expansion in 
$1/M_u$ for $M_u = M_Q$ presented in ref.~\cite{Krauss:2007bz} of 
as the former is singular in the degenerate limit. The fact that their 
expansion does not have imaginary parts can be explained by the 
freedom of phase reparametrization discussed above.

\bibliographystyle{JHEP}

\end{document}